\documentclass[fleqn,usenatbib]{mnras}

\usepackage{graphicx}	
\usepackage{amsmath}	
\usepackage{multicol}        
\usepackage{bm}		
\usepackage{listings}
\usepackage{multirow}
\usepackage{hyperref}
\usepackage{tabularray}

\DeclareRobustCommand{\Sec}[1]{Sec.~\ref{#1}}

\DeclareRobustCommand{\sec}[1]{Section~\ref{#1}}
\DeclareRobustCommand{\App}[1]{Appendix~\ref{#1}}
\DeclareRobustCommand{\app}[1]{appendix~\ref{#1}}
\DeclareRobustCommand{\Tab}[1]{Table~\ref{#1}}

\DeclareRobustCommand{\Fig}[1]{Figure~\ref{#1}}

\DeclareRobustCommand{\Eq}[1]{Eq.~(\ref{#1})}

\newcommand{\Gaia}{\textit{Gaia}}
\newcommand{\enbid}{\mbox{\textsc{EnBiD}}}
\newcommand{\enlink}{\mbox{\textsc{EnLink}}}
\newcommand{\GalaxyFlow}{\textsc{GalaxyFlow}}

\usepackage[T1]{fontenc}
\usepackage{ae,aecompl}

\usepackage{newtxtext,newtxmath}

\usepackage[shortcuts]{extdash} 

\title[GalaxyFlow]{GalaxyFlow: Upsampling Hydrodynamical Simulations for Realistic Mock Stellar Catalogs}

\author[S.H Lim et al.]{
Sung Hak Lim,$^{1}$\thanks{E-mail: sunghak.lim@rutgers.edu}
Kailash A. Raman,$^{1,2,3}$\thanks{E-mail: kailash.raman@rutgers.edu}
Matthew R.~Buckley,$^{1}$\thanks{E-mail: mbuckley@physics.rutgers.edu}
and David Shih$^{1}$\thanks{E-mail: shih@physics.rutgers.edu}
\\
$^{1}$NHETC, Department of Physics and Astronomy, Rutgers, the State University of New Jersey, Piscataway, NJ 08854, USA\\
$^{2}$Theoretical Physics Group, Lawrence Berkeley National Laboratory, Berkeley, CA 94720, USA\\
$^{3}$Berkeley Center for Theoretical Physics, University of California, Berkeley, CA 94720, USA\\
}

\pubyear{2023}

\begin{document}
\label{firstpage}
\pagerange{\pageref{firstpage}--\pageref{lastpage}}
\maketitle

\begin{abstract}

Cosmological $N$-body simulations of galaxies operate at the level of ``star particles" with a mass resolution on the scale of thousands of solar masses. 
Turning these simulations into stellar mock catalogs requires ``upsampling" the star particles into individual stars following the same phase-space density.
In this paper, we introduce two new upsampling methods. First, we describe \GalaxyFlow{}, a sophisticated upsampling method that utilizes normalizing flows to both estimate the stellar phase space density and sample from it.
Second, we improve on existing upsamplers based on adaptive kernel density estimation, using maximum likelihood estimation to fine-tune the bandwidth for such algorithms in a way that improves both the density estimation accuracy and upsampling results.
We demonstrate our upsampling techniques on a neighborhood of the Solar location in two simulated galaxies: Auriga~6 and h277. 
Both yield smooth stellar distributions that closely resemble the stellar densities seen in the \Gaia{} DR3 catalog.
Furthermore, we introduce a novel multi-model classifier test to compare the accuracy of different upsampling methods quantitatively. 
This test confirms that \GalaxyFlow{} more accurately estimates the density of the underlying star particles than methods based on kernel density estimation, at the cost of being more computationally intensive.

\end{abstract}





\section{Introduction}

Large, detailed astronomical surveys are revolutionizing our understanding of the kinematic, photometric, and spectroscopic properties of the Milky Way and its satellites. Surveys such as SDSS \citep{2000AJ....120.1579Y}, DES \citep{2005astro.ph.10346T}, and \Gaia{} \citep{2016A&A...595A...1G,2018A&A...616A...1G,2021A&A...649A...1G} have revealed aspects of the Milky Way's merger history, dark matter substructure, stellar streams, satellite dwarf galaxies, and more. As existing surveys continue and new observatories (e.g., JWST \citep{2009ASPC..418..365G}, Vera Rubin Telescope \citep{2012arXiv1211.0310L}, Grace Roman Space Telescope \citep{2015arXiv150303757S}) join them, our knowledge of the Milky Way's constituents will continue to grow tremendously.

These massive datasets require equally sophisticated simulations in order to fully understand and interpret the underlying cosmological and astrophysical parameters of the Milky Way.
Additionally, accurate simulations provide a proving ground for new analysis methods.
In particular, given \Gaia{}'s unique capability to measure proper motions, accurate and realistic simulations of \Gaia{} observations must match the kinematics of the Galaxy's stars in addition to matching the stars' spatial distribution.

Broadly speaking, two approaches exist to generate synthetic Milky Way-like galaxies with a level of detail similar to actual \Gaia{} observations. 
In the first approach, synthetic stars are sampled from an analytic model for the Milky Way. 
For example, \citet{2018PASP..130g4101R} generated a \Gaia{}-like catalog from the Besan\c{c}on model of \citet{2003A&A...409..523R} using the Galaxia code \citep{2011ApJ...730....3S}. 
Such analytic models can be closely tuned to observed properties of the Galaxy. 
Although those models are beneficial when working with \Gaia{} data, the underlying model of the Galaxy consists of completely smooth distributions with no substructure, streams, or merger history. 
For analyses that target these and other non-equilibrium properties of the Milky Way, this type of simulation may not serve all necessary purposes.

In the second approach, synthetic observations are taken of fully cosmological $N$-body simulations of Milky Way-like galaxies. 
State-of-the-art simulations consider dark matter and baryons with smoothed particle hydrodynamics, generating simulated galaxies that appear very similar to our own in many respects  — although, of course, the precise merger history and local environment of the simulated galaxies can differ from those of the Milky Way, which may have important effects on the detailed properties of the stellar kinematics.
These simulations have shed light on various interesting anomalies, such as the too-big-to-fail problem \citep{Boylan-Kolchin:2011qkt,Boylan-Kolchin:2011lmk,Brooks:2012vi,Papastergis:2014aba}, the missing satellite problem \citep{Klypin:1999uc,Moore:1999nt,Brooks:2012ah,Sawala:2012cn,Sawala:2015cdf,Wetzel:2016wro,Despali:2016meh,Garrison-Kimmel:2017zes}, and others. 
Unlike analytic models, $N$-body simulations naturally contain realistic non-equilibrium effects and kinematically-consistent substructure evolved from galaxy mergers.

However, the $N$-body simulations all operate at the level of ``star particles": objects with masses generally on the order of $10^3-10^4\,M_\odot$, each of which represents an entire population of stars. 
As a result, these simulations are not directly suitable for comparison with Milky Way stellar populations from large survey data. 
To create a realistic \Gaia{}-like survey catalog from these $N$-body simulations, star particles must be spread out in phase space via some ``upsampling'' technique. 
The goal of any such technique should be to produce a dataset of upsampled stars that follows the same kinematic phase space distribution as the original star particles, so as to preserve the self-consistent kinematics of the cosmological simulation.
In particular, a good upsampling technique should not result in kinematic artifacts or any other artificial substructure at length scales below the simulation resolution (which is ${\mathcal O}(10^2)$~pc for current state-of-the-art $N$-body simulations with baryons); instead it should be ``smooth" on scales smaller than the simulation resolution.

Several catalogs which simulate the capabilities of the \Gaia{} telescope have been constructed from $N$-body simulations of dark matter and baryons. 
\citet{2011ApJ...730....3S} and \citet{2015MNRAS.446.2274L} are early examples, as are the Aurigaia \citep{2018MNRAS.481.1726G} and the Ananke \citep{2020ApJS..246....6S} catalogs, derived from the Auriga \citep{2017MNRAS.467..179G} and the Feedback In Realistic Environments (FIRE) {\it Latte} simulations \citep{2016ApJ...827L..23W,2018MNRAS.480..800H}, respectively.

The Ananke and Aurigaia\footnote{Strictly speaking, only the \textsc{icc-mocks} catalog of Aurigaia uses \enbid{} for upsampling. The \textsc{hits-mocks} catalog of \citet{2018MNRAS.481.1726G} does not perform upsampling in position or velocity space.} catalogs use the \enbid{} algorithm \citep{enbid} to estimate the phase space density of the star particles and spread them out by adaptive kernel smoothing. 
The \enbid{} algorithm first tessellates the position/velocity space into hypercubes, each containing a single star particle, using an entropy-based criterion.
A length scale of each hypercube (such as a side length), can be used as a bandwidth for the (Gaussian) kernel smoothing.
The upsampling is accomplished by drawing samples from the kernel.

When using such kernel-based methods, it is important to choose the appropriate bandwidth for the kernel in order to reduce the bias from the smoothing.
In particular, if the bandwidth is chosen too small, {\it as was the case in both Ananke and Aurigaia}, then the upsampled distributions will be too clumpy. This lack of smoothness can be seen in plots in proper motion and Galactocentric velocity (see \Fig{fig:competitor_distributions_mu}), which reveal an unphysical blotchiness to the stars in the simulated \Gaia{} catalogs.

To improve on the poor upsampling performance of the existing Gaia mock catalogs, this paper introduces two new upsampling algorithms. Our major result is \GalaxyFlow{}, a novel synthetic star upsampling technique that employs state-of-the-art density estimation methods, specifically normalizing flows (for recent reviews and original references, see e.g.~\citealt{9089305, papamakarios2021normalizing}). Normalizing flows have previously been used to accurately estimate the phase space density of stars in a galaxy.
\citet{2020arXiv201104673G} was an early study demonstrating the idea in a Plummer sphere.
\citet{2021MNRAS.506.5721A, 2022MNRAS.511.1609N} use a mock dataset from analytic galaxy models to confirm the idea by using masked autoregressive flows \citep[MAF;][]{NIPS2017_6c1da886}, and show the estimated density can be used for recovering the gravitational potential and acceleration.
\citet{2022arXiv220501129B} applies MAF to estimate the density of stars in an $N$-body simulated galaxy and uses the density estimate to estimate the local gravitational acceleration and mass density. MAFs are used in \citet{2022MNRAS.509.5992S} as part of a model-agnostic ``anomaly detector'' to identify stellar streams in \Gaia{} data.
These successes of flows in accurately estimating stellar phase space densities in both simulated and actual astronomical datasets give us confidence that they should be effective upsamplers of galaxy simulations.

In this work, instead of using a MAF, we will use another type of normalizing flow: continuous normalizing flows \citep[CNF;][]{NEURIPS2018_69386f6b,grathwohl2019ffjord}, which have an excellent inductive bias for modeling smooth distributions. 
CNFs have been shown to be effective in recovering the densities of various analytic models \citep{2022arXiv220502244G}.
This theoretical and empirical evidence indicates that a CNF can be a good flow model for reducing the clumpiness of the upsampled dataset.
In \App{app:maf}, we will explicitly show that a CNF performs better than a MAF by our upsampling evaluation metrics.

In addition to \GalaxyFlow{}, we develop a fine-tuned \enbid{}-based upsampler.
To mitigate the inherent smoothing bias that limits the \enbid{} algorithm's ability to replicate the distribution of original star particles, we introduce a post-optimization technique to refine the bandwidths given by the original \enbid{} algorithm. 
This method treats the kernel density estimation as a Gaussian mixture model (as they have the same functional form), allowing us to adjust the bandwidths through maximum likelihood estimation of the overall bandwidth scale parameter.
This optimized \enbid{} algorithm yields a more accurate stellar distribution than both the default \enbid{} parameters and the small-bandwidth \enbid{} setup used in constructing the Aurigaia and Ananke catalogs. As we will describe, \GalaxyFlow{} and fine-tuned \enbid{} have complementary strengths and weaknesses: the latter is computationally lightweight but still suffers from some amount of oversmoothing, while the former is computationally expensive but more accurate to the underlying phase space distribution.

Both of our proposed upsampling methods can be applied to a wide range of $N$-body simulations.
As a concrete demonstration of our upsampling methods, we use the star particles in a simulated galaxy called Auriga~6 \citep{2017MNRAS.467..179G} -- the underlying $N$-body simulation used for building the Aurigaia catalog. 
Though we use the Auriga 6 simulation for specificity, nothing about \GalaxyFlow{} or the kernel method depends on the details of this simulation, and they can be applied to a wide range of $N$-body simulations. In \App{app:h277}, we show the results applied to the h277 galaxy simulation generated by the $N$-Body Shop \citep{2012ApJ...761...71Z,2012ApJ...758L..23L}.

As this paper aims to be a proof-of-concept demonstration of \GalaxyFlow{}, we will concentrate exclusively on the sampling of stellar distributions in position and velocity spaces. 
This allows a straightforward upsampling performance comparison between the normalizing flows and \enbid{} algorithms. Furthermore, we will narrow our focus to the subset of stars in a spherical region within a 3.5~kpc radius of the Sun, as going out to farther radii increases the training time dramatically. 
Also, limiting ourselves to the Solar neighborhood allows us to make qualitative comparisons to the subset of \Gaia{} DR3 stars that have full 6d positions and velocities. 
As we will see, considering stars within this window is sufficient for a detailed and definitive demonstration of the efficacy of the \GalaxyFlow{} method.

In addition to showing that the \GalaxyFlow{}- and kernel-based upsampled stellar distributions replicate the star particle distributions by eye,
we develop a novel neural network-based \emph{multi-model classifier test} to determine quantitatively which upsampling method generates a set of synthetic stars closer to the phase-space density of the original star particles.
The classifier is trained to distinguish samples from multiple upsamplers, and its output can be utilized for assigning likelihoods to each upsampler given reference dataset, which is not used for training upsamplers.
With this multi-model classifier test, we will quantitatively see that \GalaxyFlow{} and \enbid{} are separable, and the reference data is much closer to \GalaxyFlow{} than to \enbid{}. This holds true even when comparing \GalaxyFlow{} to our optimized kernel-based upsampling method.

Note that the effectiveness of the multi-model classifier test can be affected by the characteristics of the classifier used.
For example, we will employ a multilayer perceptron (MLP) as our classifier model, which is known to have a slow training issue in learning high-frequency modes \citep{pmlr-v97-rahaman19a, 10.1007/978-3-030-36708-4_22, yang2022overcoming}.
The clumpy distribution generated by the small-bandwidth \enbid{} used in the Ananke and Aurigaia catalogs has high-frequency modes.
In \App{app:justification}, we address this by setting up a small-bandwidth \enbid{} upsampler that generates stellar distributions similar to the blotchy Aurigaia catalog, and we demonstrate the validity and precision of the test when the training efficiency issue is prominent.

In future work, we will apply \GalaxyFlow{} to full hydrodynamical simulations, in order to produce complete and realistic \Gaia{} mock catalogs. A full realistic synthetic catalog must also sample stars over a range of masses and metallicities (as well as realistic dust extinction). 
As we will discuss in our conclusions, these do not present insurmountable problems, but we defer a full algorithm including these effects to future studies.

The outline of our paper is as follows.
In \sec{sec:current_status}, we describe the mock \Gaia{} catalogs and simulations we use in this paper and compare the mock catalogs to the \Gaia{} Data Release 3 (DR3) catalog.
In \sec{sec:nfs}, we introduce \GalaxyFlow{} and summarize the details of normalizing flows for the upsampling.
In \sec{sec:enbid}, the optimized \enbid{} algorithm with additional bandwidth fine-tuning used in this paper is explained.
In \sec{results}, we introduce our multi-model classifier test for comparing the performance of upsamplers, and we qualitatively and quantitatively compare the upsampled datasets from \GalaxyFlow{} and the \enbid{} algorithm for the Auriga~6 simulation.
\sec{app:computational_speed} shows the computational speed of the upsampling algorithms considered in this paper.
We conclude in \sec{conclusion} by discussing the limitations of our current upsampling and outlining future steps toward a full upsampling algorithm.
Additionally, in \App{app:maf}, we compare our upsampling results using a CNF with that of a MAF.
In \App{app:justification}, we revisit the small-bandwidth \enbid{} in order to discuss the potential limitations of the multi-model classifier test using MLP-based classifiers.
In \App{app:h277}, we demonstrate the generality of our method by applying it to a second hydrodynamical simulation, the h277 galaxy.


\section{Current status of mock \Gaia{} catalogs}
\label{sec:current_status}

To present the current status of mock \Gaia{} catalogs, in this section, we consider two different mock catalogs of \Gaia{}-like observations of a Milky Way-analogue galaxy: Ananke \citep{2020ApJS..246....6S}, based on the FIRE {\it Latte} simulations \citep{2016ApJ...827L..23W,2018MNRAS.480..800H}; and Aurigaia  \citep{2018MNRAS.481.1726G}, derived from the Auriga $N$-body simulations \citep{2017MNRAS.467..179G}.

The {\it Latte} simulations \citep{2016ApJ...827L..23W,2018MNRAS.480..800H}, used to generate the Ananke dataset, were produced using the GIZMO code \citep{2015MNRAS.450...53H}. 
The star particle mass in these simulations is initially $7070\,M_\odot$. 
The snapshot has an average particle mass of $5000\,M_\odot$, as the star particle masses can shrink over time due to the life cycle of individual stars in the assumed stellar population.

The Auriga suite of simulations was produced using the Arepo magneto-hydrodynamic code \citep{2010ARA&A..48..391S}, from comoving 100~Mpc snapshots of the EAGLE project \citep{2015MNRAS.446..521S}. 
The characteristic dark matter particle mass is $4\times 10^4\,M_\odot$, and the baryonic particle mass is $\sim 5000\,M_\odot$. 
Auriga~6, chosen due to its similarity with the Milky Way, has a total mass of $1.01\times 10^{12}\,M_\odot$ within the virial radius.

These $N$-body simulations both contain ${\cal O}(10^{7})$ star particles in their Milky Way equivalents. 
From these particles, the position and velocities of the ${\cal O}(10^{10})$ individual stars needed for a realistic simulation of the \Gaia{} catalog are drawn through an upsampling process.
As discussed in the Introduction, both Aurigaia and Ananke used the \enbid{} algorithm as the core of their upsampling method. 
This code estimates the phase-space volume taken up by each massive star particle; this information is then used to spread out the distribution of upsampled stars using a series of probability distributions (chosen to be Gaussian) centered on each star particle in the sampling step (see \sec{sec:enbid} for additional discussion of the \enbid{} algorithm).
In addition to kinematic information, the upsampling must also account for stellar masses, ages, metallicity, and observational effects such as dust extinction. 
In this work, we are primarily concerned with the kinematic sampling and will defer these other important stellar properties for later work.

\begin{figure*}
    \begin{center}
        \includegraphics[width=0.9\textwidth]{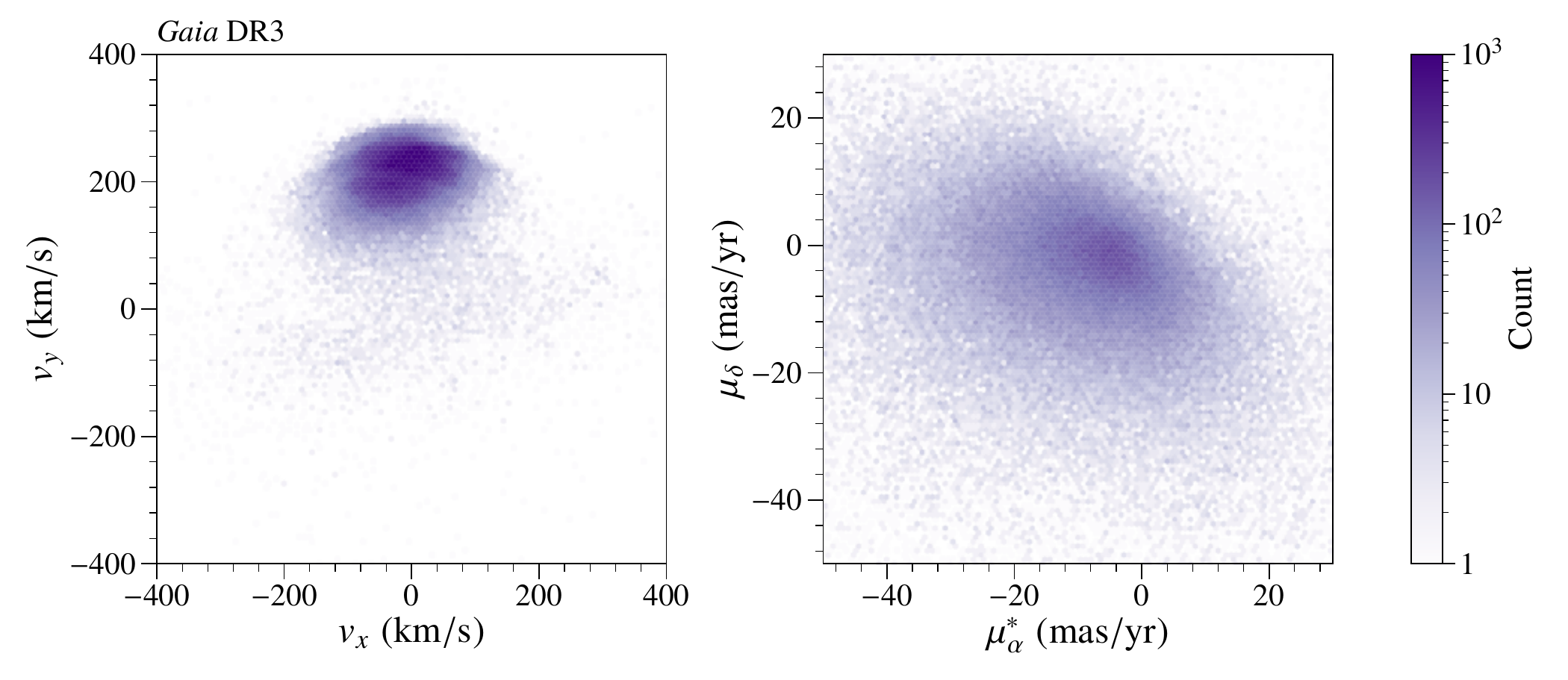} \\
        \includegraphics[width=0.9\textwidth]{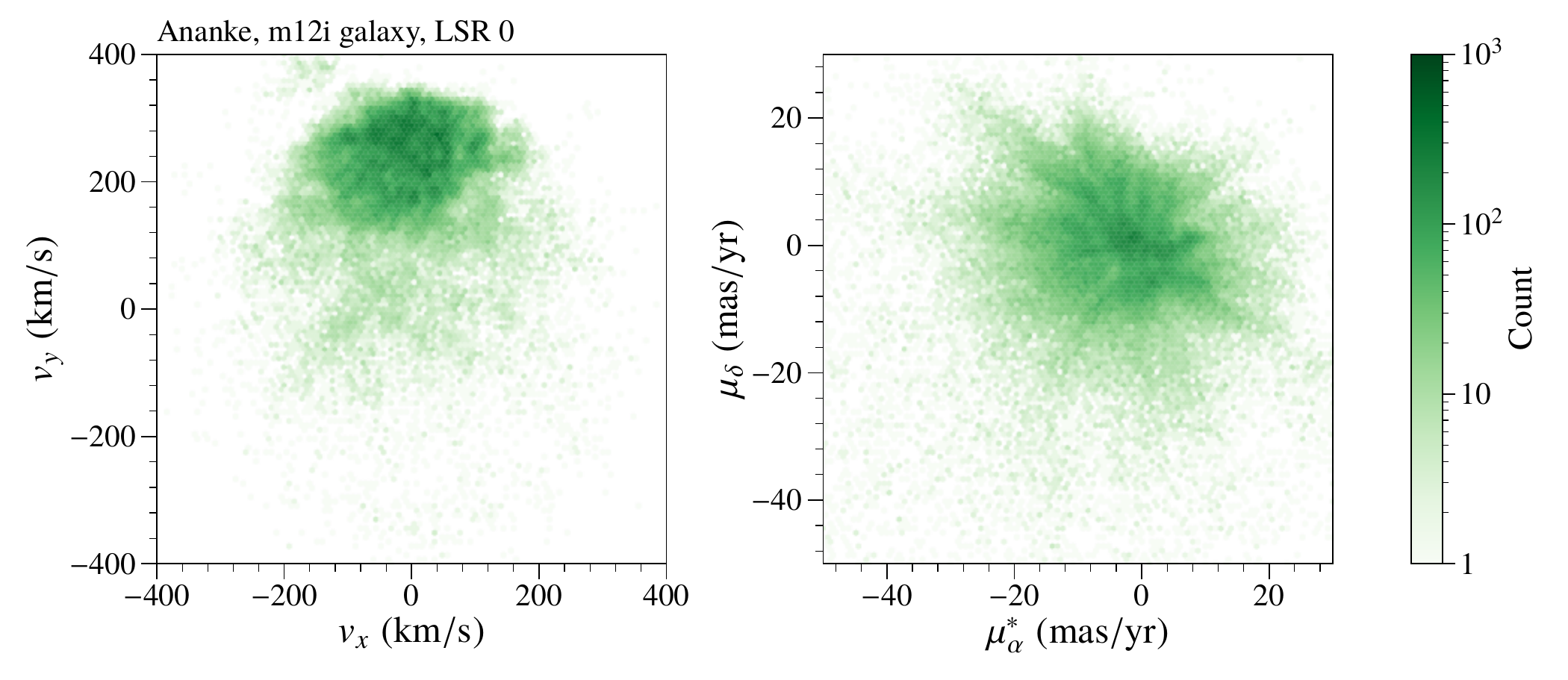} \\
        \includegraphics[width=0.9\textwidth]{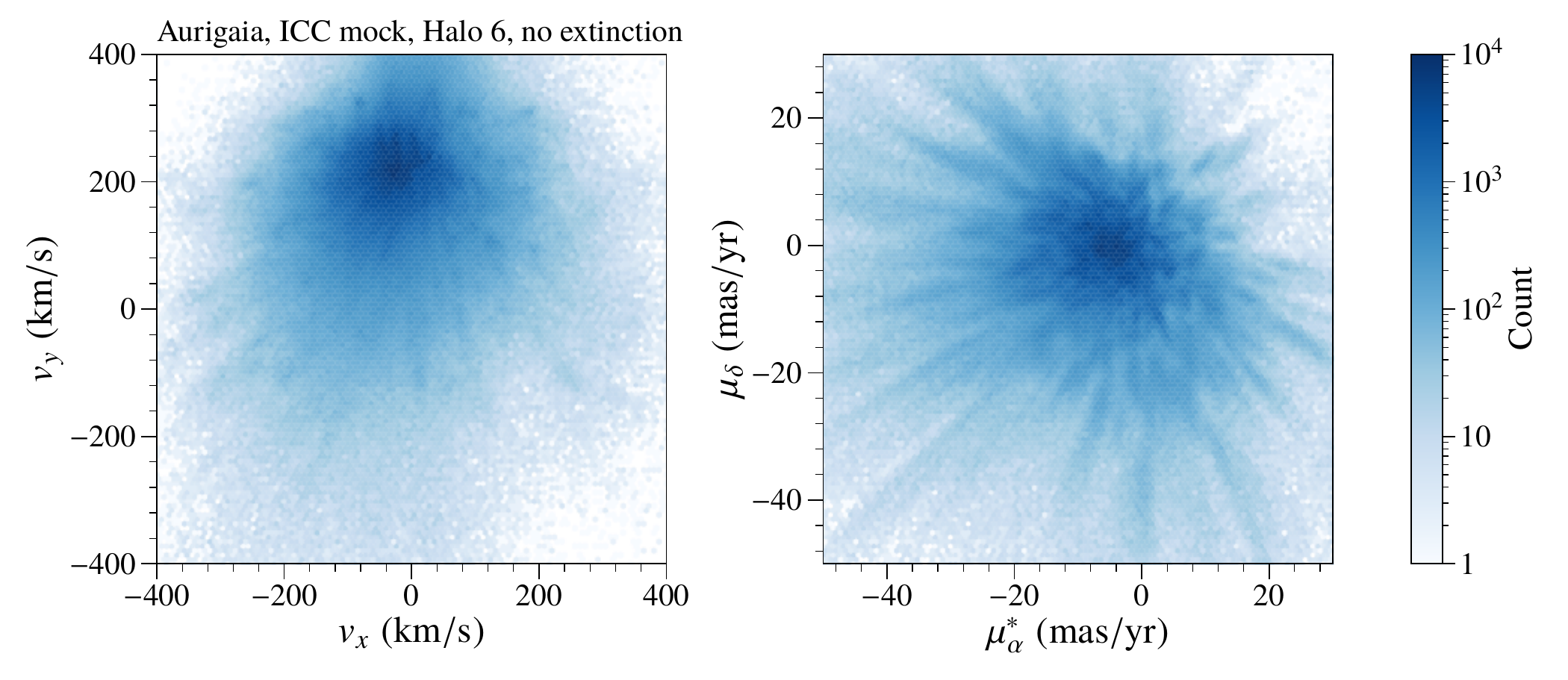}
    \end{center}
    \caption{
        Distributions of Galactocentric velocities $(v_x, v_y)$ (left column) and proper motion (right column) for all stars within $15^\circ$ of the ICRS coordinate $(\alpha,\delta) = (167.47^\circ,-4.2^\circ)$ and with measured radial velocities. 
        We additionally require that the ``observed'' parallax is more than $3\sigma$ away from $\varpi = 0$, in order to remove stars with poorly measured distances.
        Distributions are shown for real \Gaia{} DR3 data (top row), for Ananke (middle row), and for Aurigaia (bottom row).
        The \Gaia{} DR3, Ananke, and Aurigaia datasets have 139,164; 94,970; and 2,192,188 stars satisfying the selection criterion, respectively.
        Note that the Aurigaia catalog presents radial velocities of all the stars so that coverage and sample statistics are quite different from the other datasets. 
        \label{fig:competitor_distributions_mu}
    }
\end{figure*}

The kinematic sampling approaches used in Aurigaia and Ananke reproduce many features of the real \Gaia{} data, particularly in the spatial distribution of the upsampled stars. However, especially in velocity space, residual traces of the progenitor star particles remain and can be clearly seen in relatively simple plots of the distribution of stars. 
As an example, in \Fig{fig:competitor_distributions_mu}, we show -- for all stars with measured position and velocity\footnote{We select stars with measured radial velocities and with observed parallax $3\sigma$ away from the zero parallax to remove poorly measured stars. We further require positive parallax.}  in a $15^\circ$ observational window centered on the ICRS coordinate $(\alpha,\delta)=(167.47^\circ,-4.2^\circ)$ -- the distribution of proper motions on the sky as well as stellar velocities in Galactocentric coordinates. 
The center of the $15^\circ$ window was chosen to be off of the Galactic disk but is otherwise randomly selected.
These distributions are shown for real \Gaia{} DR3 data \citep{2021A&A...649A...1G} as well as the Ananke and Aurigaia simulations.

The upsampled stars exhibit clustering and streaking in the velocity and proper motion that are not present in the actual Milky Way population as seen by \Gaia{}.\footnote{Although we compare \Gaia{} DR3 to Ananke and Aurigaia DR2 mock catalogs, the difference between DR2 and DR3 cannot explain the clustering and streaking shown in the latter panels of \Fig{fig:competitor_distributions_mu}. If anything, given the considerably smaller measurement uncertainties on proper motions in DR3, comparing DR3 versions of Ananke and Aurigaia would presumably only exacerbate the discrepancies.}
We verify in \App{app:justification} that the visually-apparent clusters of simulated stars in proper motion- and velocity-space correspond to individual parent star particles in the original $N$-body simulation.

Note that the three datasets have very different numbers of stars in the same angular patch on the sky (139,164; 94,970; and 2,192,188 stars for \Gaia{}, Ananke, and Aurigaia, respectively). 
The order-of-magnitude difference in statistics between the mock catalogs is mainly because the Aurigaia catalog provides radial velocities of all the stars \citep{2018MNRAS.481.1726G}, while the Ananke catalog provides the information for the stars satisfying the magnitude and temperature limits of {\it Gaia} DR2 \citep{2020ApJS..246....6S}.
A similar order-of-magnitude difference can be seen in the total number of sources in {\it Gaia} DR3 \citep{2022arXiv220800211G} and the number of sources with radial velocities in {\it Gaia} DR2 \citep{2018A&A...616A...1G}. 
As this work is concerned only with the distribution of the stars in phase space, which can be seen regardless of these factors, we do not correct for them.

\begin{figure*}
    \centering
    \includegraphics[width=0.95\textwidth]{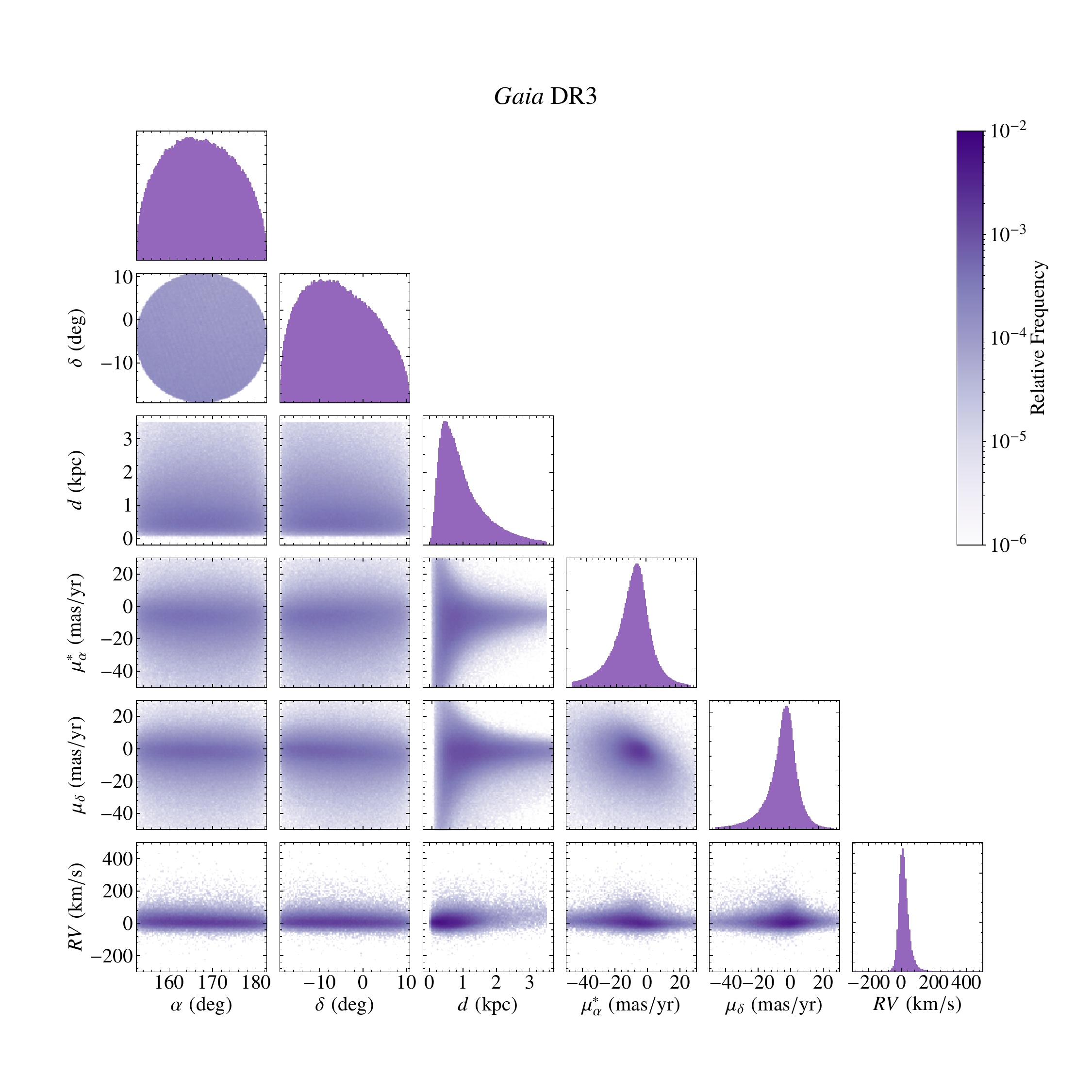}
    \caption{
        Distributions of a subset of \Gaia{} DR3 stars in ICRS coordinates.
        We select stars within $15^\circ$ of $(\alpha,\delta) = (167.47^\circ,-4.2^\circ)$ and within $3.5$~kpc of the Sun.
        We discard stars with poorly measured distances, i.e., the observed parallax must be greater than zero parallax by $3 \times$ the parallax measurement error.
        Plots on the diagonals are the marginal distribution of each component.
        The other plots are the joint distribution of two components.
        The \Gaia{} DR3 dataset has 979,185 stars satisfying these criteria.
        Only a small fraction of stars have measured radial velocity, and so the plots including radial velocity have lower statistics, with 135,105 stars.
    }
    \label{fig:gaia_icrs}
\end{figure*}

To further illustrate the comparative smoothness of \Gaia{}, we show the full phase space distribution in ICRS coordinates for the \Gaia{} DR3 stars in \Fig{fig:gaia_icrs}. The full sample (not requiring a measured radial velocity) contains 979,185 stars. 
Only a small fraction of stars have measured radial velocity, so plots of radial velocity are statistically suppressed with only 135,105 stars (all other plots use the full sample of stars). Observational effects are apparent in the distance histogram, which falls after 1~kpc instead of continuing to rise as expected for the underlying distribution of stars.
Most notably, we see no significant ``blotchiness'' in the resulting distributions.

Such spurious blotchiness often appears in kernel density estimation when the chosen bandwidth is too small, which appears to be the case in the Aurigaia and Ananke synthetic observations.
To address this failure mode, one could use a larger bandwidth for each kernel. 
When doing so, we must be careful about the smoothing bias that is inherent in kernel density estimations with finite sample sizes: the upsampled stars will respect the smeared phase-space density of star particles, not the actual phase-space density \citep{Garcia-Portugues2023}.
To upsample stars without encountering both blotchiness and smoothing bias, an alternative approach is needed. In this work, we provide two such methods: \GalaxyFlow{} using normalizing flows (\Sec{sec:nfs}), and a kernel-based method with an algorithm to select the bandwidth to avoid clumping without significant oversmoothing (\Sec{sec:enbid}).

\subsection{Benchmark Datasets for Training Upsamplers}
\label{sec:training_dataset}

Our goal is to generate upsampled stars from $N$-body simulations of Milky Way-like galaxies with stellar kinematics which display less clumping than found in the existing suites of mock \Gaia{} datasets (most notably in velocity-space) and accurately reflect the true phase-space density of star particles.
To achieve this, we train upsamplers on the kinematics of star particles from fully-cosmological $N$-body simulations containing both baryons and dark matter, and compare upsampling qualities between algorithms.
In the main body of this paper, we focus on the Auriga~6 galaxy\footnote{The simulation data for this galaxy is available from the Auriga Project at \url{https://wwwmpa.mpa-garching.mpg.de/auriga/gaiamock.html}.} \citep{2017MNRAS.467..179G}, which is the simulated galaxy used for constructing the Aurigaia mock catalog. 
The details of the simulation which produced Auriga~6 have been described in the previous section. 
In \App{app:h277}, we work with the h277 galaxy to cross-validate our general \GalaxyFlow{} method and optimized \enbid{} algorithm with a different mock dataset.

The Galactocentric coordinate system of Auriga~6 is oriented such that the galactic disk lies in the $xy$-plane, with the assumed Solar location lying in the $-x$ direction. The $z$ coordinate points out of the disk, oriented so that the local motion of disk stars is in the $+y$ direction.

The star particles within 3.5 kpc of the default Solar location of the Aurigaia catalog, $(-8.0,0,0.02)$ kpc, are selected for training our upsamplers.
The number of selected star particles in Auriga~6 is 482,412, with a maximum speed of 858.15 km/s.
Although we limit the star particles to be upsampled, this dataset is enough for comparing the upsampling performance of various algorithms. 
Training upsamplers over a full galaxy will be addressed in future work.

The Aurigaia mock catalog aims to model the real \Gaia{} data accurately and so must sample stellar populations from isochrones, creating synthetic stars with a range of magnitudes and colors. 
After this sampling, realistic observational effects must be applied. 
In this work, we are interested only in the sampling of the kinematic phase space: comparing our new techniques using normalizing flows and the optimized \enbid{}-based approach.
For this comparison, the full simulation of synthetic stars (including isochrones and observational effects) is unnecessary, so we omit these steps.

\section{\GalaxyFlow{}}
\label{sec:nfs}

\subsection{Normalizing Flows}
\label{flows}

\GalaxyFlow{} is based on a type of generative model called normalizing flows.
Normalizing flows are a class of neural networks that learn a transformation from a simple base probability distribution to a function representing the unknown probability distribution from which the training data was drawn. 
A fully trained normalizing flow allows for both density estimation of a dataset as well as sample generation from the approximated density distribution.
A summary of the idea is as follows.

Let $\vec{z}$ and $\vec{y}$ be $D$-dimensional random variables with the base distribution and training data distribution, respectively.
The standard normal distribution is conventionally used as the base distribution because of its tractability. 
Normalizing flows will learn a bijective transformation $g$ between $\vec{z}$ and $\vec{y}$:
\begin{equation}
    \vec{y} = g(\vec{z}).
\end{equation}
Once the network has learned this transformation, it is straightforward to upsample the dataset using normalizing flows: randomly select samples from a multidimensional standard normal distribution and then transform them to synthetic data in the training space using the above formula.

By composing a chain of simple, nonlinear bijective functions $g_i$:
\begin{equation}
    g = g_N \circ g_{N-1} \circ \cdots \circ g_1.
\end{equation}
one can build a highly expressive family of bijections $g$ that are capable of modeling non-trivial distributions.
For computational efficiency, we require that inverse and Jacobian determinants of those simple bijections are easy to compute.

Training normalizing flows are based on maximum likelihood estimation.
If the inverse transformation of $g$ and the Jacobian determinant $\left|{d\vec{z}}/{d\vec{y}}\right|$ are easy to compute, the change of variable formula can be used to compute the probability density function of $\vec{y}$:
\begin{equation}
    p(\vec{y}) 
    = 
    p(\vec{z}) \cdot \left| \frac{d\vec{z}}{d\vec{y}} \right|
    =
    p(g^{-1}(\vec{y})) \cdot 
    \left| \frac{d
    g^{-1}(\vec{y})}{d\vec{y}} \right|.
\end{equation}
By maximizing $\log p(\vec y)$ (summed over the training data) with respect to the parameters of $g$, the normalizing flow can be fit to the data and learn the underlying probability density.

In addition, normalizing flows are able to learn conditional probabilities if the transformation
$g$ is conditioned. This capability is especially useful in modeling the phase space density $f(\vec{r},\vec{v})$, as the full 6-dimensional density need not be learned all at once. Instead,
we borrowed the well-performing setup from our previous work \citep{2022arXiv220501129B} and used two separate flows learning position density $p(\vec{r})$ and velocity density conditioned on position, $p(\vec{v} | \vec{r})$.\footnote{In \citet{2022arXiv220501129B}, this decomposition was necessary in order to efficiently evaluate velocity integrals at a given position appearing while solving the Boltzmann equation and Gauss's law. For the upsampling problem, we need only the phase space density, so this decomposition is not strictly necessary.} The phase space density $f(\vec{r},\vec{v})$ is modeled by their product as follows.
\begin{equation}
    f(\vec{r},\vec{v}) = p(\vec{r}) p(\vec{v} | \vec{r}).
    \label{eqn:prob_decomposition}
\end{equation}
Note that we do not need to train $p(\vec{r})$ and $p(\vec{v} | \vec{r})$ simultaneously.
We will train $p(\vec{r})$ first and then $p(\vec{v} | \vec{r})$. 

\subsection{Continuous Normalizing Flows}

Within the general class of normalizing flows, we have to choose an optimal implementation for smoothly upsampling star particles. 
Continuous normalizing flows \citep[CNF;][]{NEURIPS2018_69386f6b,grathwohl2019ffjord} are a good candidate with an inductive bias suitable for this problem because the transformation smoothly deforms the base distribution to the target distribution.

More specifically, CNF learns the following infinitesimal transformation,
\begin{eqnarray}
    g_t 
    & : & 
    \vec{y} \rightarrow \vec{y} + F(\vec{y},t) \cdot  dt,
    \\
    g_t^{-1} 
    & : & 
    \vec{y} \rightarrow \vec{y} - F(\vec{y},t) \cdot  dt.
\end{eqnarray}
Here, the function $F$ is a neural network representing the derivative ${d \vec{y}}/{dt}$ of the trajectory of transformed variables at a latent time $t$. 
The full chain of transformations is the integral of this infinitesimal transformation, and it is described by a neural ordinary differential equation \citep[neural ODE;][]{NEURIPS2018_69386f6b},
\begin{equation}
    \frac{d}{dt}\vec{y}(t) = F(\vec{y}(t),t).
\end{equation}
Note that if $dt$ is finite, the transformation $g_t$ is essentially a residual block at a given time, $\vec{y} \rightarrow \vec{y} + \mathcal{F}(\vec{y})$, where $\mathcal{F}$ is the difference between the inputs and outputs of the transformation. 
Therefore, the neural ODE is considered as a generalization of residual networks for normalizing flows \citep{haber2018learning,pmlr-v80-lu18d,Haber_2018,ruthotto2020deep}, and the parameter $t$ takes the role of the flow index in the chain.

The Jacobian determinant of the transformation can be obtained by solving the following form of the Fokker-Planck equation with zero diffusion \citep{NEURIPS2018_69386f6b}, describing the time evolution of log probability $\log p(\vec{y}(t); t)$ along the trajectory $\vec{y}(t)$ at time $t$:
\begin{equation}
\frac{d}{d t} \log p(\vec{y}(t); t ) = - \mathrm{Tr} \left[ \frac{\partial F}{\partial \vec{y}(t)}\right].
\label{eqn:cnf_jaco}
\end{equation}
The trace computation of this equation is often a bottleneck during the training, so Hutchinson's trace estimator \citep{grathwohl2019ffjord} can be used to speed up this step.
In our case, the cost of evaluating the trace is manageable since we train CNFs for 3D densities; we explicitly evaluate the trace during the training.

Our CNF is implemented using \textsc{Torchdyn} \citep{politorchdyn} with backend \textsc{PyTorch} \citep{NEURIPS2019_bdbca288}.
The neural ODEs are solved by the Runge-Kutta method of order 4 for both forward and backward directions.
For the backward direction, we could numerically invert the forward calculation for the exact invertibility of the flows \citep{pmlr-v97-behrmann19a,NEURIPS2019_5d0d5594} but use the numerical solver again for fast calculation.
We will see that this approximate inverse transform is still good enough to model a good upsampler.
We set the distribution at $t=0$ to be the base distribution and that at $t=1$ be the target distribution, and the latent time interval is divided into 20 steps.
The derivative function $F$ is modeled by a multilayer perceptron (MLP) consisting of 4 hidden layers with 32 nodes each.
GELU activations \citep{2016arXiv160608415H} are used in order to model a smooth transformation.
For our conditional CNF for modeling $p(\vec{v}|\vec{r})$, the conditioning variables $\vec{r}$ are provided as extra inputs to $F$.

\subsection{Training Normalizing Flows for Upsampling Star Particles}
\label{sec:preprocessing}

We preprocess the position and velocity of the star particles in order to standardize the inputs and separate the sharp boundaries at the edge of the training volume. 
The CNFs model a smooth transformation in the latent space, so we preprocess to avoid discontinuities.
The position vectors $\vec{r}$ are preprocessed as follows in order to construct normalizing flows on an open ball in Euclidean space \citep{2022arXiv220501129B}:
\begin{enumerate}
    \item 
    {\it Centering and scaling:} We scale the positional coordinates to transform the 3.5~kpc radius sphere centered on the Sun into a unit ball by the following linear transformation:
    \begin{equation}
        \vec{r} \rightarrow \frac{\vec{r} - \vec{r}_{\sun}}{r_{\mathrm{max}} \cdot c},
    \end{equation}
    where $r_{\mathrm{max}}$ is the radius of the selection window, $\vec{r}_{\odot}$ is the location of the Sun in the galaxy, and $c=1.000001$ is a small constant to avoid putting stars exactly on the boundary.
    \item
    {\it Radial transformation:} We then expand the unit ball to cover Euclidean space, i.e.,
    \begin{equation}
        \vec{r} \rightarrow \frac{\vec{r}}{|\vec{r}|} \tanh^{-1} \left| \vec{r} \right|.
    \end{equation}
    This transformation is designed to have well-defined derivatives at the origin.
    \item \label{enum:standardization}
    {\it Standardization:} We finally rescale the position coordinates by the width of their distributions:
    \begin{equation}
        r^a \rightarrow \frac{r^a - {\mu}_{r^a}}{ \sigma_{r^a}}, \quad a = 1,2,3 ,
    \end{equation}
    where ${\mu}_{r^a}$ and $\sigma_{r^a}$ are the mean and standard deviations of $r^a$ in the training dataset, respectively.
\end{enumerate}
The velocity distribution does not have a sharp boundary at the edge of the observational sphere, so preprocessing the velocity vectors consists only of standardizing the inputs (Step \ref{enum:standardization} above).

Two normalizing flows regressing $p(\vec{x})$ and $p(\vec{v}|\vec{x})$ in the preprocessed coordinate are trained as follows.
The loss function is defined as the negative log-likelihood of the flow evaluated on the data.
We use an ADAM optimizer \citep{2014arXiv1412.6980K} with a learning rate of $10^{-3}$ to minimize the loss function.
We randomly select 20\% of the training samples as a validation dataset.
The remaining training samples are randomly split into 10 mini-batches for each epoch of training.
We stop the training when the validation loss has not improved over 50 epochs.
The trained network is further refined by restarting training with a reduced learning rate of $10^{-4}$.
We train 10 instances of normalizing flows initialized with different random number seeds, and we ensemble average to improve the performance further.

We generate position and velocity samples using the normalizing flows as follows.
\begin{enumerate}
    \item
    Sample Gaussian random numbers from the base distributions.
    \item
    Use the trained normalizing flows to transform the random numbers into position and velocity vectors in the preprocessed space.
    \item
    Use the preprocessors to transform the vectors in the preprocessed space to the position and velocity vectors.
    \item
    We remove outliers by discarding samples with distances from the Sun larger than 3.5~kpc, or with speeds exceeding the maximum speed of the training sample. 
\end{enumerate}

\section{\enbid{} Density Estimation}
\label{sec:enbid}

We compare the phase space distributions generated by the normalizing flows to the distribution generated using the \enbid{} algorithm \citep{enbid} — the adaptive kernel density estimation used as the base of the upsampling method for constructing the existing state-of-art \Gaia{} mock catalogs. 
In this paper, we use an \enbid{} algorithm as implemented in \enlink{} 0.1.0 \citep{enlink, enlink.repo}.
Here, we briefly summarize how the \enbid{} algorithm determines the bandwidth of the kernels and upsamples stars afterward. We then describe a method to select an optimal bandwidth for the kernel, preventing the clumpiness that characterises existing synthetic stellar populations that were constructed using kernel-based methods.

Given a set of star particles in phase space, \enbid{} first partitions the phase space into disjoint boxes, each containing exactly one particle.
Each box represents a volume with a unit probability of finding a particle, $1/N$, where $N$ is the total number of particles.
The boxes are constructed using the following entropy-based criterion:
\begin{enumerate}
    \item
    Start with a box containing all the star particles.
    \item
    For a given box containing $N_p$ star particles, calculate the Shannon entropy along each axis. 
    The entropy is estimated by modeling the probability density as a histogram of $N_p$ equal-sized bins. 
    \item
    Select the axis with minimum entropy.
    \item
    Divide the box in two by dividing the selected axis at the midpoint of the nearest upper and lower neighbors of the mean value of the corresponding component.
    \item
    Repeat the above splitting until all the boxes contain only one particle.
\end{enumerate}
Dividing the axis with minimum entropy prioritizes splitting along axes with clustered structures.
This box splitting rule results in small box volumes in regions of high density.
The inverse of a box's volume can be used as a proxy for the phase space density within that box.
Each star particle $i$ can then be upsampled by drawing from (Gaussian) kernels with smoothing bandwidths $\vec h_{(i)}$ determined by hypercube length scales. These per-star-particle smoothing bandwidths can be viewed as the ultimate output of the \enbid{} algorithm.

We note that Aurigaia \citep{10.1093/mnras/stu2257,2018MNRAS.481.1726G}  and Ananke \citep{2020ApJS..246....6S} appear to have incorrectly extracted the $\vec h_{(i)}$ parameters from \enbid{}, leading to upsampling bandwidths that are much too small. This explains why their upsampling star distributions are so blotchy in phase space.

\subsection{Fine-Tuning Bandwidths using a Mixture Model}

The clumpiness visible in existing upsampled datasets is not a fundamental limitation of \enbid{}, and can be alleviated if the bandwidth is appropriately set. Here we describe an algorithm which further refines the \enbid{} bandwidths to enhance the upsampling quality. The results give kinematic distributions which are smooth, though we will show that some oversmoothing remains and the samples drawn from the CNF are quantatively more similar to the underlying data.

In order to fine-tune the bandwidths, we construct a likelihood model of bandwidth scaling parameter $c_h$ by considering the kernel density estimation with bandwidths scaled by $c_h$ as a Gaussian mixture model.
The resulting likelihood function of $c_h$ is as follows.
\begin{equation}
    f(\vec{w}; c_h) 
    =
    \frac{1}{N} \sum_{i=1}^N K(\vec{w}| \vec{w}_{(i)}, c_h \vec{h}_{(i)}), \quad \vec{w} = (\vec{r}, \vec{v}).
    \label{eqn:kde}
\end{equation}
Here, $N$ is the number of samples in the kernel density estimation, 
$K(\vec{w}|\vec{\mu},\vec{\sigma})$ denotes a Gaussian kernel with mean $\vec{\mu}$ and bandwidth $\vec{\sigma}$, 
and $\vec{w}_{(i)}$ corresponds to the position and velocity vector of $i$-th training sample.
The scale parameter $c_h$ adjusts the overall scale of kernel bandwidths $\vec{h}_{(i)}$ given by the \enbid{} algorithm using the number of nearest neighbors, $k=64$.
We note that this modeling can be used for fine-tuning bandwidths given by any adaptive kernel density estimations.

To find an optimal $c_h$, we minimize the negative log-likelihood of the validation dataset.
Given that this optimization involves a single parameter, we employ the Newton-Raphson method for its efficiency and fast convergence.

\begin{figure}
    \begin{center}
       \includegraphics[width=0.49\textwidth]{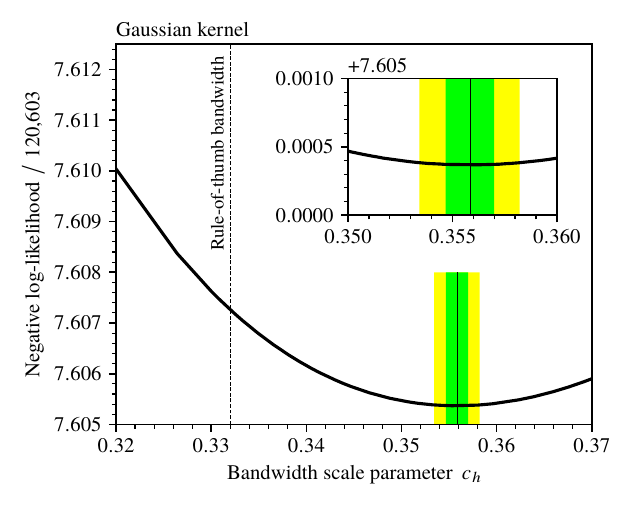}

    \end{center}
    \caption{
        The negative log-likelihood function of the scale parameter $c_h$.
        The kernel density is constructed using half of the star particles in the training dataset, and the log-likelihood function is evaluated using the other half. 
        Each half of this training dataset contains 120,603 star particles.
        The best-fit value of $c_h$ is 0.356. Green and yellow bands represent $1\sigma$ and $2\sigma$ confidence intervals, respectively.
        The Black dashed vertical line is the scale parameter estimated from the ratio between the rule-of-thumb bandwidths of the spherical Epanechnikov kernel and the kernel used in the plot.
    }
    \label{fig:bandwidth_scale}
\end{figure}

We present the bandwidth scale optimization results in \Fig{fig:bandwidth_scale}.
For the upsampling using half of the star particles as the validation set, the optimal scale parameter $c_h$ is $0.356\pm0.001$.\footnote{The optimal scale parameter for the spherical Epanechnikov kernel (the default kernel of \enlink{} package) is $1.080\pm0.004$, giving a ratio of these coefficients of $0.330\pm0.014$. 
This is perfectly consistent with the rule-of-thumb bandwidth ratio between the spherical Epanechnikov and Gaussian kernel, 0.332, derived in \app{app:rot_bandwidth}.}
We use this scaling parameter to correct the \enbid{} bandwidth for each particle. This optimized choice allows us to avoid the clumpiness of the existing upsampled catalogs while also not greatly oversmoothing the phase space density.

To further improve the resulting dataset, we perform \enbid{} upsampling only after applying the preprocessing outlined in \sec{sec:preprocessing}.
This preprocessing also helps mitigate the boundary bias inherent in kernel density estimation, which occurs when a kernel is placed near the data boundary and extends beyond it. This causes the density to be underestimated near the edges and generates outliers.
The preprocessing steps push the boundary to infinity and so greatly reduce this form of bias.

Compared to the \GalaxyFlow{} upsampler, our \enbid{} upsampler using a fine-tuned kernel is much more computationally inexpensive (see \Sec{app:computational_speed}). However, as we will discuss in the next section, though the resulting upsampled stellar populations appear smooth by eye,  \GalaxyFlow{} outperforms our kernel-based method by our full classifier metrics.

\section{Results}
\label{results}

\subsection{Qualitative Comparisons}
\label{motion_plots}

As \Gaia{} observations cover the full sky, and studies of Galactic kinematics are concerned with the correlations of stellar velocities with their location in position-space, we compare the results of upsampling the parent star particles in the same observational window as in \Fig{fig:competitor_distributions_mu}.
For Auriga~6, this patch contains 3,015 star particles.
We upsample the star particles by a factor of 500 to have statistics comparable to that of the Aurigaia catalog: 2,192,188 stars.

\begin{figure*}
    \begin{center}
      	\includegraphics[height=0.400\textwidth]{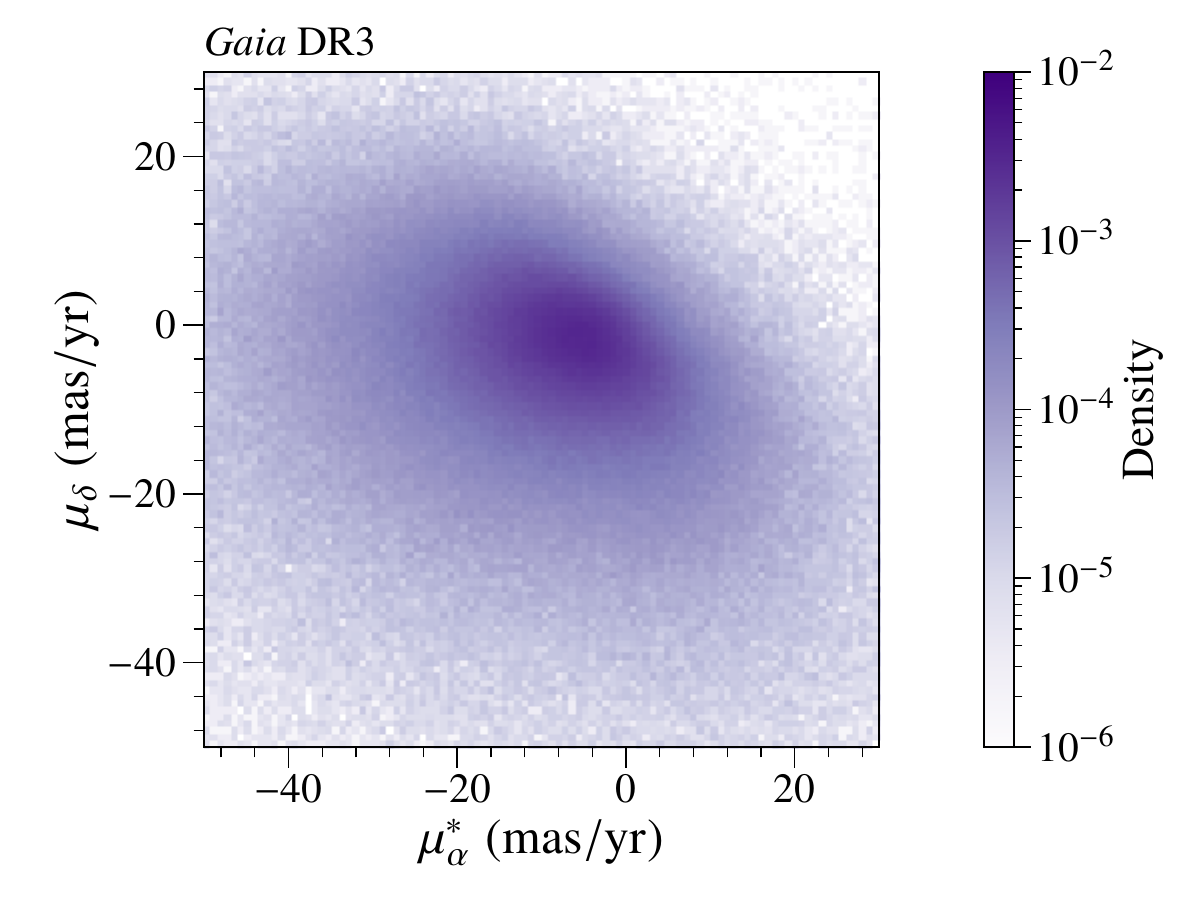}
    \end{center}
    \begin{center}
        \includegraphics[height=0.400\textwidth, trim={0 0 5cm 0}, clip]{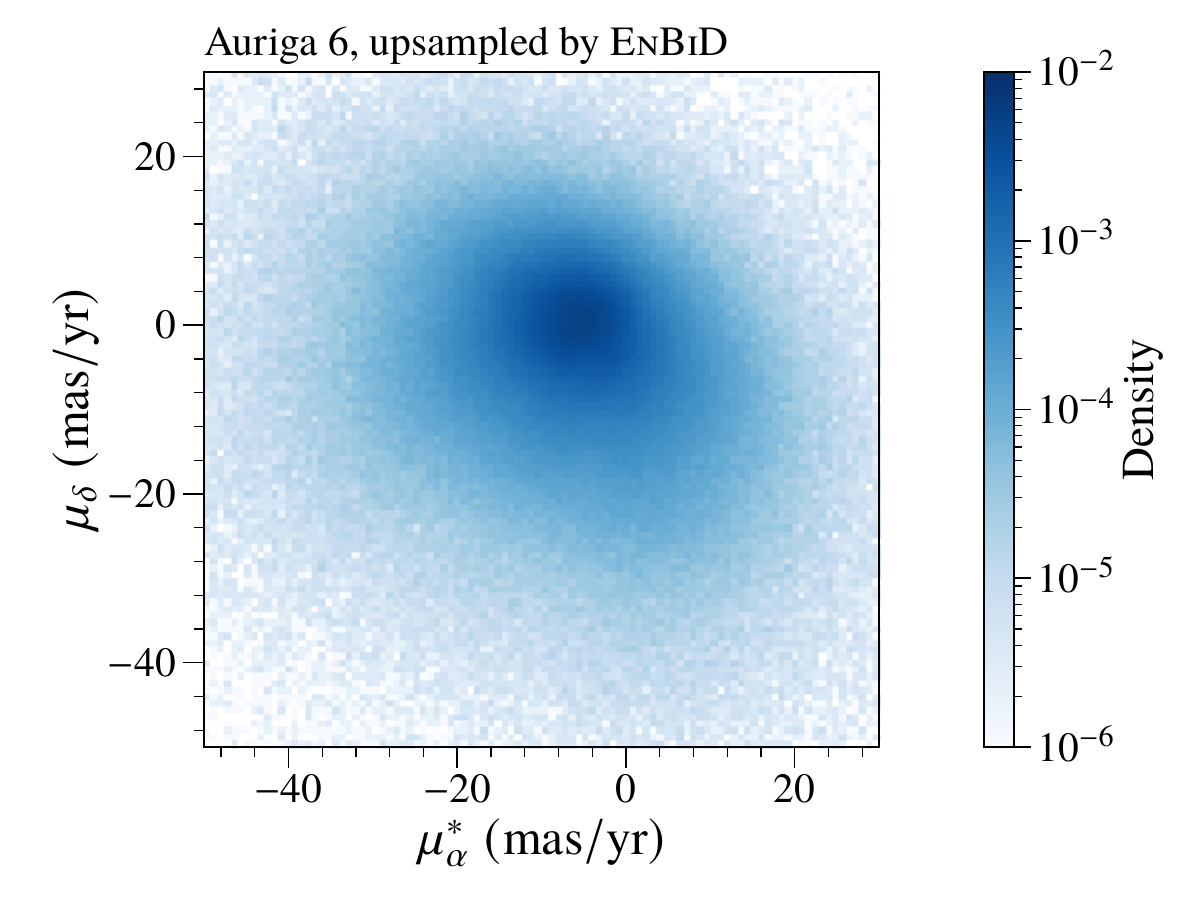}
       	\includegraphics[height=0.400\textwidth]{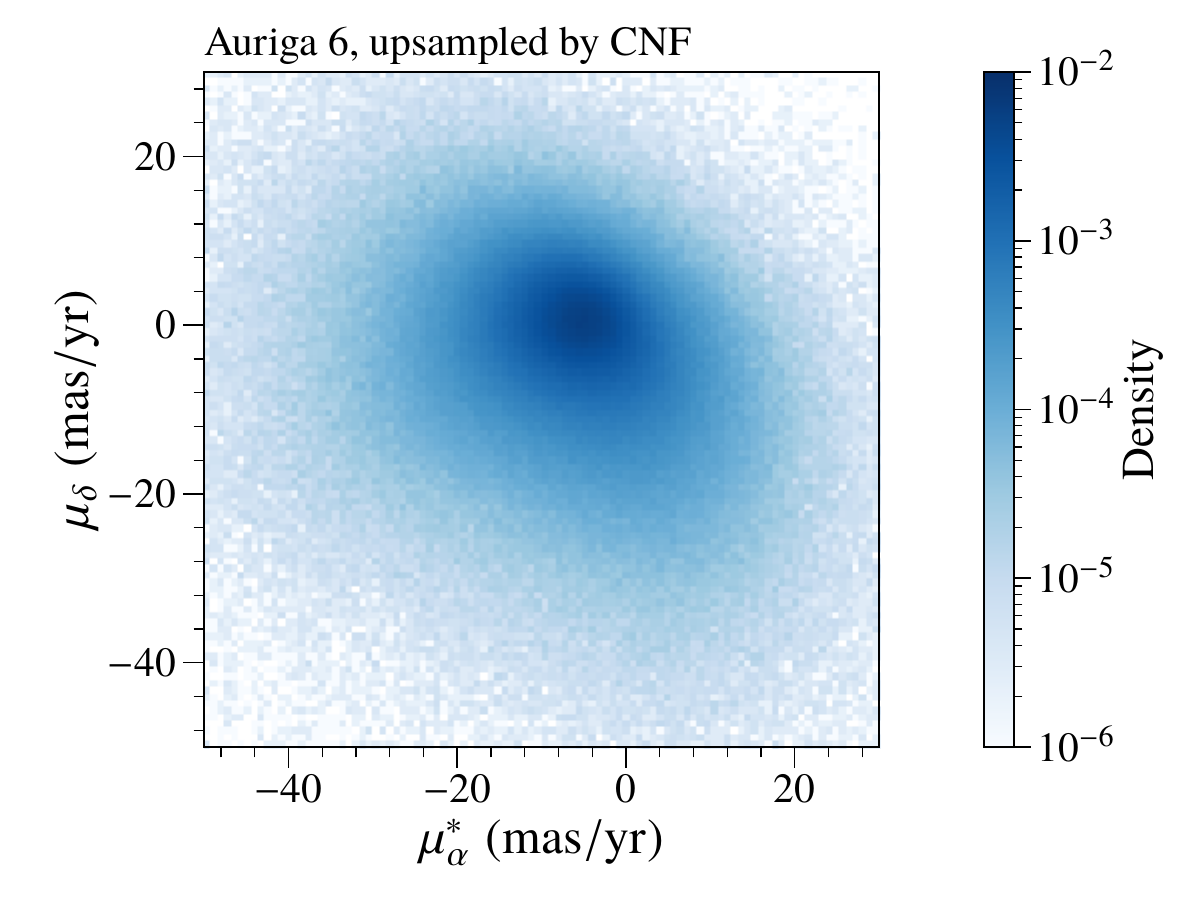}
    \end{center}
    \caption{
        Distributions of proper motion for stars in \Gaia{} DR3 (center) and upsampled Auriga~6 stars by \enbid{} using an optimized bandwidth (lower left) and CNF (lower right) (lower right) within $15^\circ$ of the ICRS coordinate $(\alpha,\delta) = (167.47^\circ,-4.2^\circ)$ with distance from the assumed Solar location within $3.5$~kpc.
        For \Gaia{} DR3, we additionally require that the ``observed'' parallax is more than $3\sigma$ away from $\varpi = 0$, in order to remove poorly measured stars.
        After the selection, the \Gaia{} DR3 dataset has 979,185 stars.
    }
    \label{fig:auriga_upsampling}
\end{figure*}

In \Fig{fig:auriga_upsampling}, we plot the proper motion for stars upsampled in the patch for the \Gaia{} DR3 catalog and from the Auriga~6 simulations using both \enbid{} and the CNF. 
The bottom right panel of \Fig{fig:auriga_upsampling} shows the proper motion of our \enbid{}-upsampled dataset.
The optimized bandwidth is sufficiently larger than the inter-particle spacing so that the clumpy substructures visible in \Fig{fig:competitor_distributions_mu} are no longer there.
The proper motion distribution is as smooth as \Gaia{} DR3 plot in the center.

The bottom left panel of \Fig{fig:auriga_upsampling} shows the proper motion of the CNF-upsampled dataset.
The CNF upsampling also generates a very smooth distribution as CNF models a smooth transformation from the standard Gaussian to the data distribution.
Within this patch, both optimized \enbid{} and CNF apparently generate high-quality proper motion distributions.

\begin{figure*}
    \centering
    \includegraphics[width=0.95\textwidth]{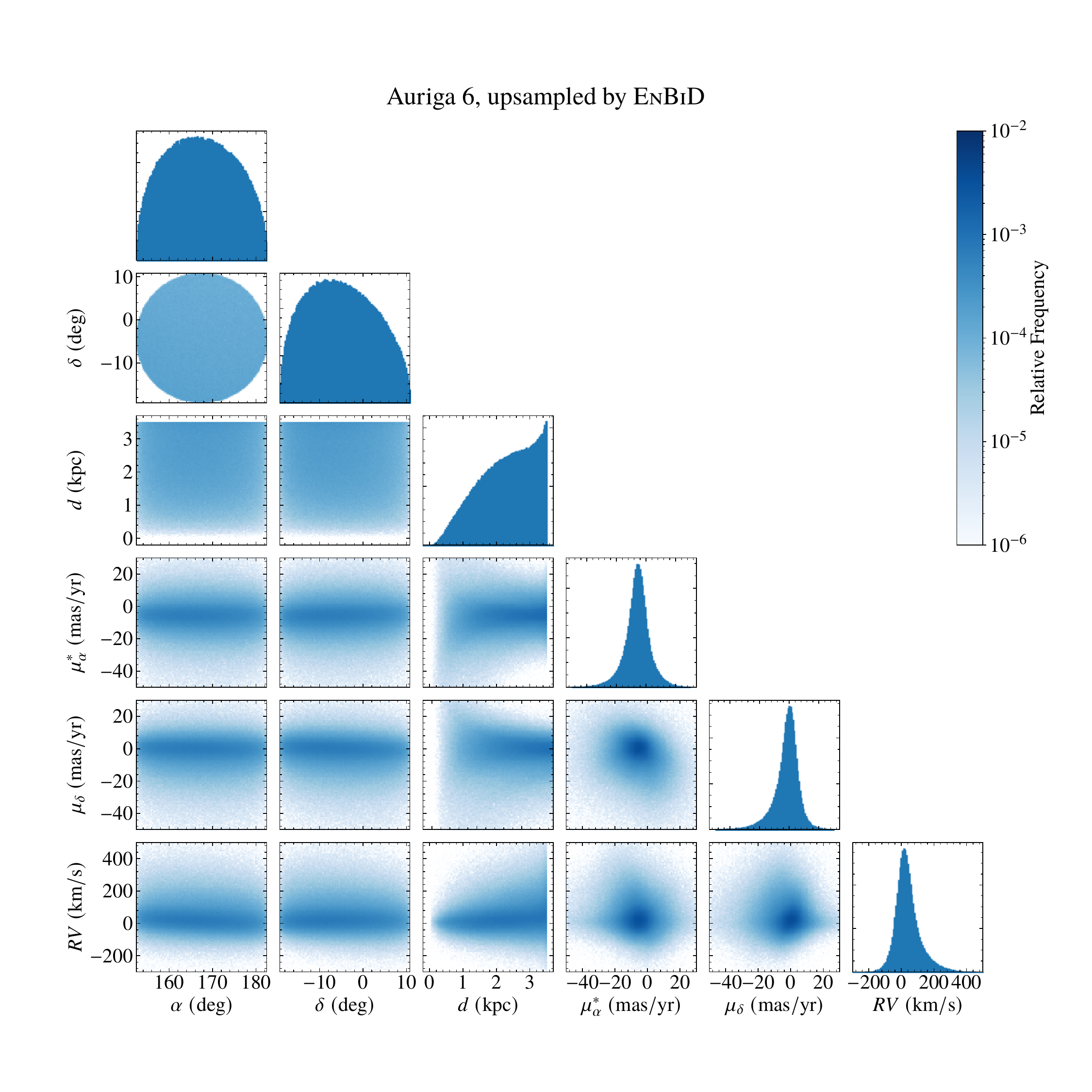}
    \caption{
        Distributions of \enbid{}-upsampled Auriga~6 stars in the ICRS coordinate.
        We select stars within $15^\circ$ from $(\alpha,\delta) = (167.47^\circ,-4.2^\circ)$ and with distances from the assumed Solar location less than $3.5$~kpc.
    }
    \label{fig:enbid_icrs}
\end{figure*}

\begin{figure*}
    \centering
    \includegraphics[width=0.95\textwidth]{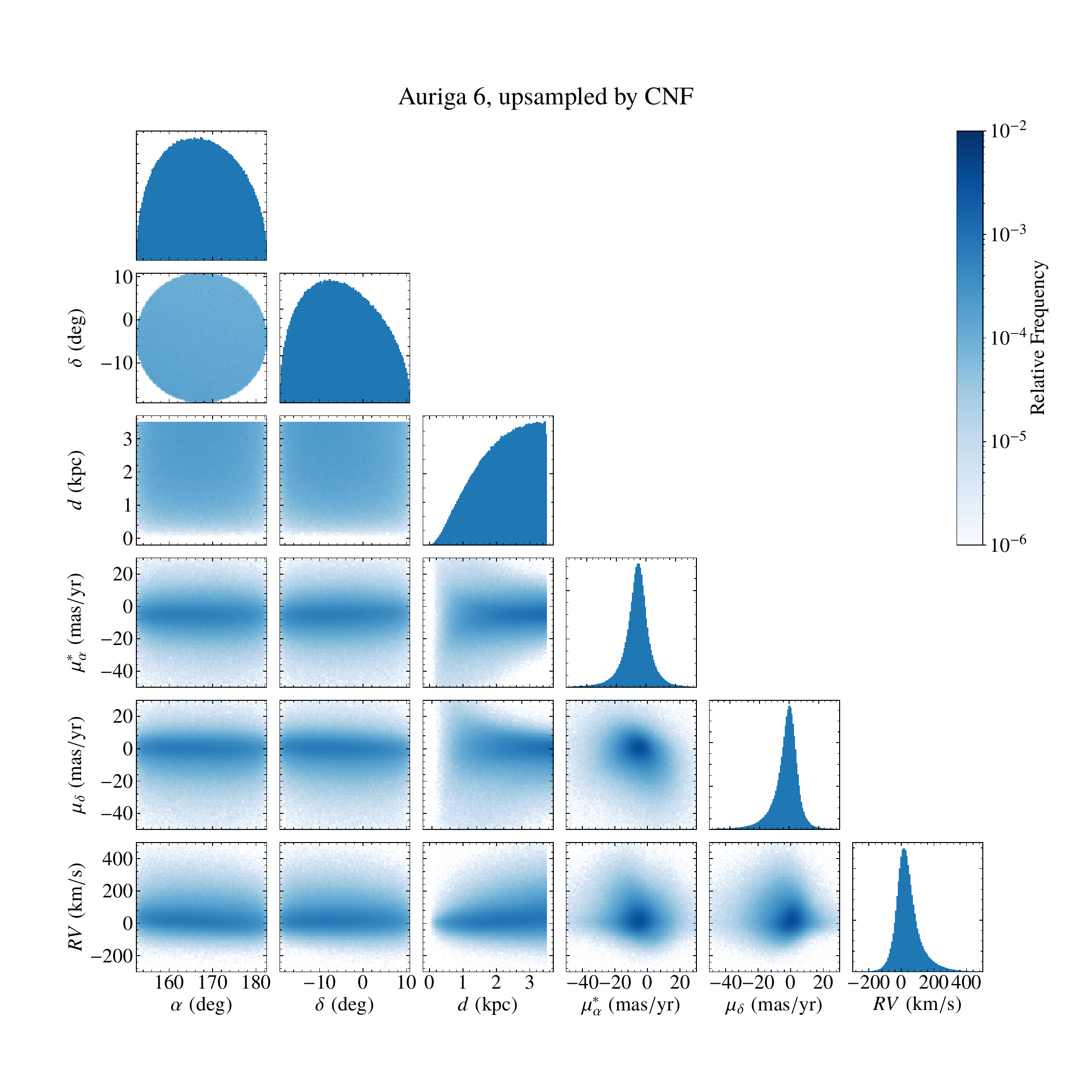}
    \caption{
        Distributions of CNF-upsampled Auriga~6 stars in the ICRS coordinate.
        We select stars within $15^\circ$ from $(\alpha,\delta) = (167.47^\circ,-4.2^\circ)$ and with distance from the assumed Solar location less than $3.5$~kpc.
    }
    \label{fig:cnf_icrs}
\end{figure*}

In \Fig{fig:enbid_icrs} and \Fig{fig:cnf_icrs}, we provide the full six-dimensional phase space (in the ICRS coordinates) of stars upsampled in the patch by our bandwidth-optimized \enbid{} algorithm and the CNF-based \GalaxyFlow{}, respectively.\footnote{We note that these plots lack observational effects, so they differ from the \Gaia{} plots, especially in the distance distribution.
The corresponding \Gaia{} distribution is a convolution of the true distances with observational effects such as observed magnitude limits.
The far-away stars are dimmer and more difficult to observe, and so we see fewer distant stars compared to the upsampled star distributions.
Nevertheless, those observational effects do not wash out the upsampler dependence, so comparing the simulation truth distributions is sufficient for our purpose.
}
Overall, both sets of upsampled distributions look similar and sufficiently smooth in all the position and velocity components, but the two upsampling methods noticeably differ in the distance histogram. 
The main reason for the sudden rise of distance distribution in \enbid{}-upsampling is due to over-smoothing in kernel density estimation.

\begin{figure*}
    \begin{center}
        \includegraphics[width=0.31\textwidth]{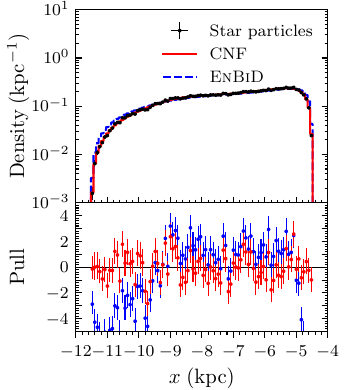}
        \includegraphics[width=0.31\textwidth]{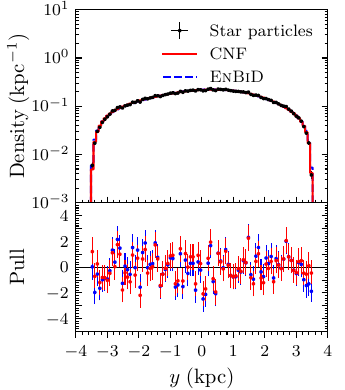}
        \includegraphics[width=0.31\textwidth]{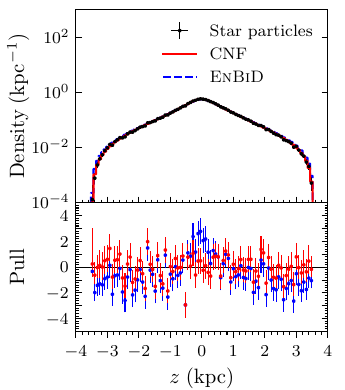}
    \end{center}
    \begin{center}
        \includegraphics[width=0.31\textwidth]{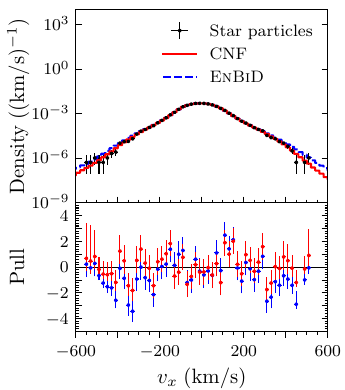}
        \includegraphics[width=0.31\textwidth]{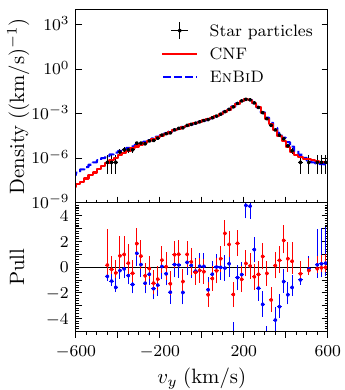}
        \includegraphics[width=0.31\textwidth]{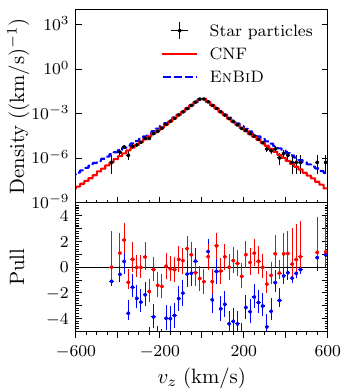}
    \end{center}
    \caption{
    Normalized histograms of (top) position components and (bottom) velocity components for selected stars in Auriga 6.
    The red lines are the histograms for synthetic stars sampled from CNF, and the blue lines are for the stars from \enbid{} using an optimized bandwidth.
    The error bars are the $1\sigma$ statistical uncertainty.
    Below the main plots, we show the pull distributions, i.e., the difference between the histograms of star particles and upsampled stars, normalized by the $1\sigma$ statistical uncertainty.
    The red and blue markers are for CNF and \enbid{}, respectively.
    }
    \label{fig:enlink_upsample_xyz}
\end{figure*}

The deviation in the \enbid{} upsampled stellar distribution from the star particle distribution is also visible in the whole dataset, 
as shown in the histograms of galactocentric cartesian coordinate components in \Fig{fig:enlink_upsample_xyz}.
The bias is particularly prominent in the areas where the density changes rapidly and at the boundary, such as at the edges of $x$-histogram and the peak of $v_y$ histogram.
Again, this deviation is due to over-smoothing.
The smearing effect of the Gaussian kernel introduces a bias proportional to the second derivative of the density, leading to sizable deviations around density peaks.

In contrast, CNF aims to estimate the true density directly through maximum likelihood estimation.
The histograms for coordinate components upsampled by CNF, depicted in \Fig{fig:enlink_upsample_xyz}, show smooth and less biased stellar distributions. 
This qualitative comparison highlights the effectiveness of CNF in replicating the star particle density without the bias introduced by kernel smearing, offering a more accurate model for upsampling.

\subsection{Two-Sample Classifier Metric}
\label{sec:classifier_test_two_sample}

While by-eye comparisons of the outputs of the upsampling methods are useful, they are necessarily qualitative, as well as limited to one or two dimensions. 
For a more quantitative, full-phase-space test of the quality of the generative models, we turn to a novel neural classifier test.

The ideal metric to judge the quality of a generative model would be the Neyman-Pearson (NP) optimal binary classifier between reference and generated samples. 
The classifier output would be monotonic with the likelihood ratio between the reference and generated distributions. 
If the two follow the same probability distribution (i.e.~$p_{\mathrm{ref}}=p_{\mathrm{gen}}$), the NP-optimal classifier would be maximally confused, i.e.~the area under the receiver operating characteristic curve (AUC) of the classifier would be 0.5. 
Conversely, the higher the AUC, the worse the generative model is at modeling the underlying probability distribution of the data.

This ``ultimate classifier metric" was recently used by \citet{2021arXiv210605285K} in a Large Hadron Collider (LHC) application to demonstrate the fidelity of a fast calorimeter simulation based on masked autoregressive flows trained on GEANT4 reference showers. 
Using this metric, the authors successfully demonstrate that normalizing flows are better at generating synthetic calorimeter responses than generative adversarial networks.
Hence, we may also try this classifier test to determine a better star particle upsampler.

However, this ``ultimate classifier metric" has many limitations and drawbacks. Chief among these is that it is only strictly reliable when the classifier is close to optimal. We find this is definitely not the case here because there are simply not enough reference star particles to train a good classifier. In the LHC case, we can always run more GEANT4 simulations to generate more reference data, but here we are limited to the fixed number of star particles in the Auriga simulation. 
As a result, the classifier training starts to overfit in the very early phase of training, and we end up with AUCs close to 0.5 (0.555 and 0.504 for \enbid{} and CNF, respectively). 
This small AUC difference does not agree with what we found with our qualitative, by-eye comparisons of \enbid{} and the reference Auriga simulation data, and indicates the classifier is not optimal.

Another issue with the ``ultimate classifier metric" is that the AUC can be misleading, in the sense that a perfectly suitable generative model could be nevertheless 100\% separable from the reference data due to some physically irrelevant ``tells" (leading to an AUC of 1). Also, the AUC has rather limited expressiveness as a measure of quality in that it is bounded between 0 and 1. 
For example, two generative models could both have an AUC of 1, but one could be much better than the other, and it would be impossible to tell using the ``ultimate classifier metric."

For these reasons, in the next subsection, we introduce a new classifier-based evaluation score to judge the relative merits of different upsampling generative models, which circumvents the problems with the ``ultimate classifier metric."

\subsection{Multi-Model Classifier Metric}
\label{sec:classifier_test}

We dub our new evaluation score the ``multi-model classifier metric". This score gives up on judging the absolute quality of a generative model with respect to the reference data, and instead focuses on judging the quality of a generative model {\it relative to a collection of other generative models}. In other words, we seek to rank a collection of generative models in terms of which one describes the reference data best.

Given a collection of $N$ generative models, we will first train an $N$-class classifier to distinguish between upsamplings of the same reference data by all the different generative models.
Since the training is over upsampled data, we are not limited by training statistics, we are able to learn the difference between these two generative models and verify that they are not equivalent.
After this training, we utilize the classifier to assign upsampling performance scores to each upsampler and identify the best upsampler as the one whose probability distribution is closest to the reference dataset.
Again, this test cannot tell us the absolute quality of a generative model, but it can tell us which generative model is most similar to the reference data.

We identify the better upsampler by using the average of log-posteriors of each upsampler evaluated on the reference dataset. 
The log-posterior of a generative model $C$ is defined as follows,
\begin{equation}
    \mathrm{LP}(C) 
    =
    \frac{1}{N_{\mathrm{ref}}} \sum_{i=1}^{N_{\mathrm{ref}}} \log p(C\,|\,\vec{r}_{(i)},\vec{v}_{(i)}),
\end{equation}
where $p(C \,|\,\vec{r}_{(i)},\vec{v}_{(i)})$ is the classifier output for model $C$, evaluated on $i$\=/th sample $(\vec{r}_{(i)},\vec{v}_{(i)})$ in the reference dataset.

One advantage of using the log-posterior as our metric is that the difference of posteriors of two upsamplers has a clear probabilistic interpretation.
The log-posterior $\mathrm{LP}(C)$ of an upsampler $C$ differs only from the log-likelihood $\mathrm{LL}(C)$ (the log-likelihood that the dataset is drawn from the probability distribution of upsampler $C$) by a constant which is the same when evaluated over both upsamplers:
\begin{eqnarray}
    \label{eqn:loglikelihood}
    \mathrm{LL}(C) 
    & = &
    \frac{1}{N_{\mathrm{ref}}} \sum_{i=1}^{N_{\mathrm{ref}}} \log p(\,\vec{r}_{(i)},\vec{v}_{(i)} \,|\, C), 
    \\
    \label{eqn:loglikelihood_and_logposterior}
    & = &
    \mathrm{LP}(C)
    +
    \frac{1}{N_{\mathrm{ref}}}  \sum_{i=1}^{N_{\mathrm{ref}}} \log \sum_{C'} p(\vec{r}_{(i)},\vec{v}_{(i)}\,|\, C').
\end{eqnarray}
Note that we train classifiers using a uniform prior of $p(C) = 1/N$, and the prior dependence of \Eq{eqn:loglikelihood_and_logposterior} is canceled out.
Therefore, the difference between the log-posteriors of two different models $C_1$ and $C_2$ can be considered as the log-likelihood ratio from one upsampler from the other, making these numbers suitable for systematically comparing the performance of upsamplers.

Another advantage of the log-posterior score is that it is not bounded and remains informative even when comparing generative models which may be separable from the reference data. In that case, the AUC becomes uninformative ($\mathrm{AUC}\approx1$ for both upsamplers), whereas the better model should still be apparent using the log-posterior because our method is based on likelihood ratios for determining the model closer to the reference dataset.

In the rest of this subsection, we will describe our implementation of the multi-model classifier test for $N=2$ models: \enbid{} and CNF. 
In \App{app:maf}, we will show an example of the multi-model classifier test with $N=3$, by including another high-performing upsampler based on the MAF-family of normalizing flows.

We will use a classifier modeled by MLP taking the position and velocity vectors $(\vec{r}, \vec{v})$ as inputs.
The classifier is trained by minimizing the categorial cross-entropy loss function so that the network will return the posterior probability $p(C|\vec{r}, \vec{v})$ of the generative model $C$.
The MLP consists of four hidden layers with [2048, 1024, 128, 128] nodes each. 
We use LeakyReLU activations \citep{maas2013rectifier} in order to avoid dying ReLU problems that may appear when the datasets are too similar.

As the classifier is for testing the generalization performance of the upsamplers, we perform the classifier training and hypothesis testing as follows:
\begin{enumerate}
    \item
    \label{enum:mmc_dataset}
    Split the selected star particles into training and test datasets for the upsampling. 
    We will use a split ratio of 1:1. 
    Each split contains 241,206 star particles. 
    \item
    Train upsamplers using the training dataset of star particles.
    \item
    For each upsampler, generate upsampled datasets with $241,206 \times 200$ samples for training the classifiers.
    Samples outside of the selection window are removed.
    \item
    Train a classifier for comparing the two upsampled datasets.
    \item
    Evaluate the log-posterior on the test star particles to determine the better upsampler.
\end{enumerate}
Half of the upsampled dataset is used as a validation dataset for the classifier. 
The remaining training datasets are split into 1000 batches for each epoch during the training.
The other setups for the classifier training are the same as those for training normalizing flows.
We stop the training when the validation loss has not improved over 100 epochs.

\begin{figure}
    \includegraphics[width=0.495\textwidth]{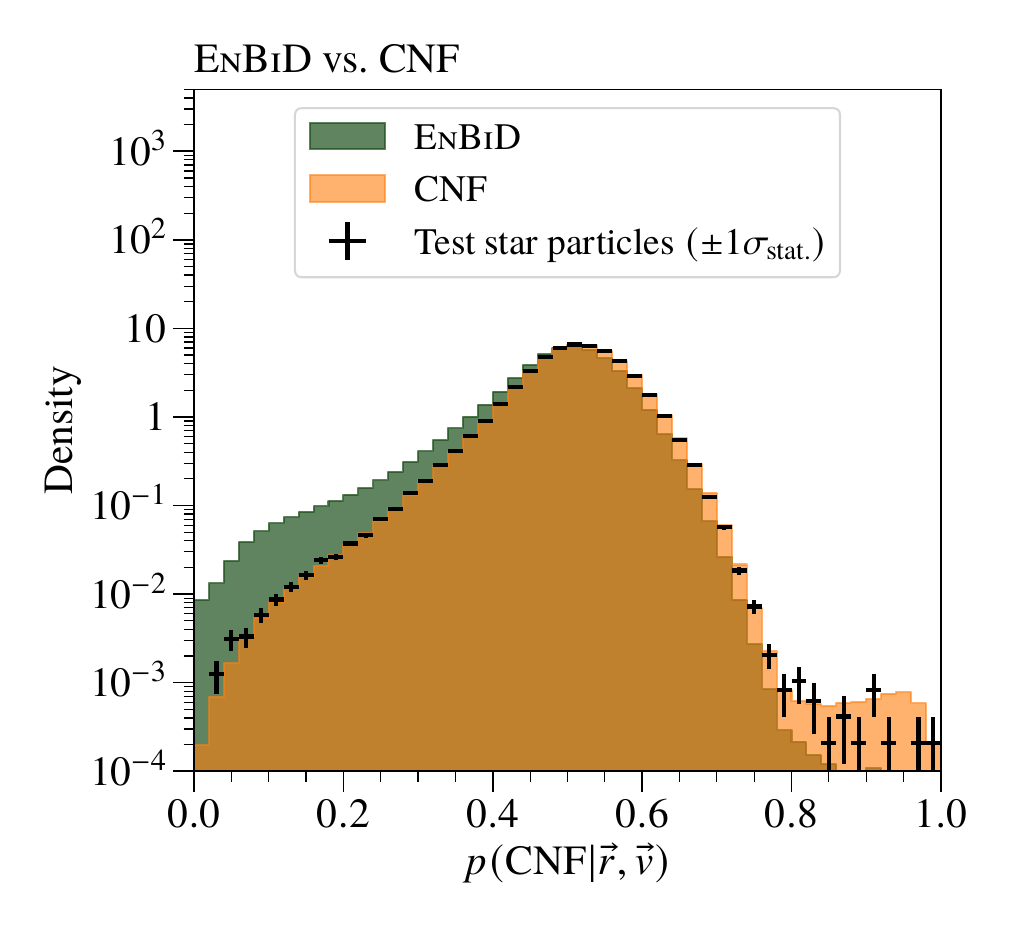}
    \caption{
        \enbid{} (green) vs. CNF (orange) classifier output distributions.
        Black markers are the classifier output histogram of star particles from the test set, which is not used in training upsamplers.
        Their vertical error bars are the $1\sigma$ statistical uncertainties.
    }
    \label{fig:classifier_enbid_cnf}
\end{figure}

\begin{table}
    \caption{
    Log-posteriors from the multi-model classifier test for \enbid{} vs.~CNF.
    Large (i.e.~less negative) log-posterior scores indicate better upsampling performance.
    We also show the $1\sigma$ standard deviations of the scores obtained by repeating ten times the upsampling and classifier training.
    The bold values are the scores of the best model determined by the test.
    }
    \begin{center}
        \begin{tblr}{cc}
            \hline
            Classifier: & \enbid{} vs.~CNF
            \\
            \hline
            Upsampler & $\mathrm{LP}(\mathrm{upsampler})$
            \\
            \hline
            \enbid{}  & $-0.724 \pm 0.003$  \\
            CNF       & $\mathbf{-0.683 \pm 0.003}$          \\
            \hline
        \end{tblr}
    \end{center}
\label{table:log_posterior_enbid_cnf}
\end{table}

First, let's compare the datasets upsampled by \enbid{} and CNF.
As shown in \Fig{fig:enlink_upsample_xyz}, the \enbid{}-estimated stellar distribution exhibits biases near edges and peaks.
In contrast, the CNF-upsampled dataset does not have such biases.
The classifier can notice this difference, leading to distinct classifier output distributions for the two upsampled datasets, as illustrated in \Fig{fig:classifier_enbid_cnf}.
The AUC of this classifier is 0.577.
The two upsampled datasets are distinguishable not only visually but also statistically through the classifier.

Next, we utilize the classifier to determine which upsampler yields a stellar distribution closest to the original star particles.
Given the biases observed in \enbid{}-upsampled stars, we expect that the classifier would more likely identify test star particles as CNF-upsampled stars.
Indeed, the classifier output for the test star particles aligns more closely with the CNF-upsampled dataset, as demonstrated in \Fig{fig:classifier_enbid_cnf}.
This suggests that the classifier perceives the test star particles as more similar to those upsampled by CNF, which were trained on a separate set of star particles.

The results of the multi-model classifier test are detailed in Table \ref{table:log_posterior_enbid_cnf}, with the log-posterior for CNF being larger than that for \enbid{}. 
Specifically, the test gives a log-posterior of $-0.683\pm0.003$ for CNF, while \enbid{}'s log-posterior is $-0.724\pm0.003$. 
The uncertainties associated with the log-posterior scores are determined by conducting the test ten times using different random seeds, covering both the statistical uncertainty of repeated upsampling and the systematic uncertainty of network initialization.

These test results quantitatively confirm that our \GalaxyFlow{} algorithm based on CNF produces stars that are more kinematically consistent with the underlying star particle distribution. 
To further validate our findings, we also cross-validate the comparison of \enbid{} and CNF using an additional mock dataset from the simulated galaxy h277. 
The results align with our results using Auriga 6, summarized in \App{app:h277}.

\section{Computational Speed Comparison}
\label{app:computational_speed}

In this section, we compare the computational speed of the upsamplers.
\Tab{tab:computing speed} shows the training and sampling speed of the upsamplers\footnote{In \Tab{tab:computing speed}, we additionally show a result of masked autoregressive flow (MAF), which is a simpler normalizing flow also tested for learning the phase space density of star particles \citep{2022arXiv220501129B}. 
We only show its computational performance here.
Although MAF is faster than CNF, the upsampling quality of MAF is slightly worse than CNF. 
Details of the test results are presented in \App{app:maf}.}, and \enbid{} is the fastest. 
For \enbid{} training, once bandwidth calculation is finished, further optimization does not take much time.
We may use either the rule-of-thumb bandwidth scaling to bypass the training or optimize the bandwidth scale parameter $c_h$, whose training only takes a few epochs.
\enbid{} sampling is also fast since we only need to select one of the components and draw a Gaussian random number from there.

For CNF, the training takes more time because of solving a multi-parameter optimization problem, which requires multiple epochs to find the best-fit solution.
Sampling is also slow because Gaussian random numbers have to be transformed to the data distribution using neural networks.

Although CNF shows the best upsampling performance, its training and upsampling speed are the slowest among the upsamplers. This might not be an issue since the upsampling needs to be performed only once per galaxy simulation, so the computational expense is a one-time cost. 
Nevertheless, it could be interesting to explore recent innovations in speeding up the CNF training from the machine learning community, for example, the probabilistic flow of score-based diffusion models \citep{song2021scorebased} and conditional flow matching \citep{lipman2023flow,tong2023conditional}.
We will leave the upsampling using these new machine learning architectures for future studies.

\begin{table}
    \caption{
        \label{tab:computing speed}
        Training and sampling times of \enbid{}, MAF, and CNF. 
        Training for \enbid{} bandwidth scaling, MAF, CNF was performed using Tesla P100; Quadro M6000; GeForce 2080 Ti, which have FP32 computing performances of 9.3 TFLOPS, 6.8 TFLOPS, and 13.5 TFLOPS, respectively.
        For \enbid{}, the training time is divided into two components: the first number is the bandwidth calculation time on a CPU (Intel Xeon E5-2630), and the second number is the bandwidth optimization time on the Tesla P100.
        To measure the upsampling time, we generate 241,206,000 samples on the Tesla P100.
        The bold values are the training and upsampling time of the fastest upsampler.
    }
    \begin{center}
        \begin{tabular}{crr}
            \hline
            Upsampler & training time & upsampling time 
            \\
            \hline
            \enbid{} & \textbf{(99 + 554) s} & \textbf{33 s}\\
            MAF        & 4158 s  & 523 s\\
            CNF        & 55090 s & 3239 s\\
            \hline
        \end{tabular}
    \end{center}
\end{table}

\section{Conclusions}
\label{conclusion}

We introduced two improved star particle upsamplers. Our primary result is \GalaxyFlow{}, based on continuous normalizing flows, which can be used for generating realistic \Gaia{} mock catalogs from $N$-body simulated galaxies.
\GalaxyFlow{} more accurately captures the phase space distribution of the original star particles than the existing state-of-the-art \enbid{} algorithm, which is an adaptive kernel density estimation specifically designed for upsampling star particles. We also showed a bandwidth-optimization for kernel-based upsampling using the \enbid{} algorithm. Our optimization avoids clumping in phase space, while mitigating oversmoothing.

In order to quantitatively compare the accuracy of \GalaxyFlow{} and our kernel-based method, we developed a novel multi-model classifier test that compares the accuracy of generative models.
This test statistically confirms \GalaxyFlow{}'s enhanced upsampling accuracy compared to the optimized \enbid{} in two examples of upsampling stars in a neighborhood of the Sun in two simulated galaxies: Auriga~6 and h277.
The improved accuracy of \GalaxyFlow{} upsampling makes the algorithm suitable for extending the kinematic features of star particles to an upsampled stellar distribution. The bandwidth-optimized \enbid{} upsampler also provides visually smooth distributions at a lower computational cost.

Although \GalaxyFlow{} shows excellent upsampling performance, there remains room for improvement.
Our upsampling only assumes the smoothness of the flow model and does not impose any physical constraints, such as the equations of motion or any specific physics-based analytic density models. 
The multi-model classifier test guarantees the kinematic consistency of upsampled data only up to the scale of the star particles: 
regression artifacts may appear on smaller scales.

To make the learned distribution at a smaller resolution scale more reliable, we may impose physical constraints.
For example, we may require that the flows satisfy the Boltzmann equation by considering additional loss functions penalizing the solution incompatible with the equation of motion \citep{RAISSI2019686}.
As our normalizing flows are estimating well the phase space density, snapshots of star particles at different times can be used for calculating the time derivative of the phase space density.
As the simulations provide a truth-level gravitational field, the Boltzmann equation can be utilized for refining the quality of the density estimation for upsampling.

In this first paper, we applied our upsampler only on a small neighborhood around the Sun, and did not simulate the range of stellar masses or dust extinction that a realistic synthetic star population as seen by \Gaia{} or other observatories must display.

While the second issue of star properties can be straightforwardly solved using existing methods, the generalized performance for upsampling over all the star particles in the simulated galaxy may differ from the results of this paper because of the soft topological constraints of normalizing flows.
When the training dataset contains vastly different topological structures (for example, multimodality) compared to the base distribution, normalizing flows sometimes experience a mild difficulty in modeling the distribution as it is modeling a continuous bijective transformation preserving the topological structure of the data.
Therefore, simple normalizing flows with a Gaussian base distribution may experience difficulties in learning galactic substructures, such as galactic streams and satellite galaxies, on top of the smooth distribution of stars.
Nevertheless, those difficulties can be mitigated by introducing more expressive flows or base distribution with multimodality.
As one potential failure mode, if the flow-based upsamplers are connecting two separated clusters, then due to a lack of expressive power, the upsampler may generate a fake unphysical stream that does not exist in the simulation.
Alternatively, if the upsampler fails to model a real smooth stream and leaves a gap, this may give us false positive signals on a stream colliding with unknown objects.
In future studies, explicit checks on the output of the normalizing flows must be carefully done to ensure that they preserve these structures and improvements developed to deal with these or other failure modes.

\section*{Acknowledgements}

We thank Alyson Brooks for helpful discussions and assistance with the simulated galaxy h277 from the $N$-Body shop \citep{2012ApJ...761...71Z,2012ApJ...758L..23L}.

This work was supported by the US Department of Energy under grant DE-SC0010008.

This work presents results from the European Space Agency (ESA) space mission Gaia. Gaia data are being processed by the Gaia Data Processing and Analysis Consortium (DPAC). Funding for the DPAC is provided by national institutions, in particular, the institutions participating in the Gaia MultiLateral Agreement (MLA). The Gaia mission website is \url{https://www.cosmos.esa.int/gaia}. The Gaia archive website is \url{https://archives.esac.esa.int/gaia}.

The authors acknowledge the Office of Advanced Research Computing (OARC) at Rutgers, The State University of New Jersey for providing access to the Amarel cluster and associated research computing resources that have contributed to the results reported here. URL: \url{https://oarc.rutgers.edu}

\section*{Data Availability}
This work uses the \Gaia{} DR3 dataset hosted in FlatHUB repository, at \url{{https://flathub.flatironinstitute.org/gaiadr3}}. 
The Ananke m12i galaxy dataset is available in the FIRE project website, at \url{{https://fire.northwestern.edu/ananke/}}. 
We use its synthetic Gaia survey hosted in FlatHUB repository. 
The Auriga 6 galaxy dataset and its synthetic survey are available on the Auriga Project website at \url{{https://wwwmpa.mpa-garching.mpg.de/auriga/gaiamock.html}}. 
The h277 galaxy dataset is available in the N-Body Shop, at \url{{https://nbody.shop/data.html}}. 
The processed datasets underlying this article will be shared on reasonable request to the corresponding author.

\bibliographystyle{mnras}
\bibliography{galaxyflow}

\appendix

\section{Comparing MAF and CNF}
\label{app:maf}

In this appendix, we compare the upsampling performance of CNFs to another model of normalizing flows -- masked autoregressive flows \citep[MAFs;][]{NIPS2017_6c1da886}.
MAFs have been studied for the phase space density estimation of the stars in mock datasets based on analytic models and $N$-body simulated galaxies, and it shows good performance on estimating derived quantities such as gravitational acceleration and mass density \citep{2021MNRAS.506.5721A, 2022MNRAS.511.1609N, 2022arXiv220501129B}. These successes suggest that the density estimation of a MAF on the $N$-body simulated stars performs well.
Nevertheless, we find the CNF has a better inductive bias when modeling smooth distributions. It is instructive to repeat the classifier test between MAF and CNF to determine which normalizing flow is better for the phase space density estimation required for upsampling.
We use the same network architecture for the MAFs in this work as in \citet{2022arXiv220501129B}.

\begin{figure}
    \begin{center}
      	\includegraphics[height=0.350\textwidth]{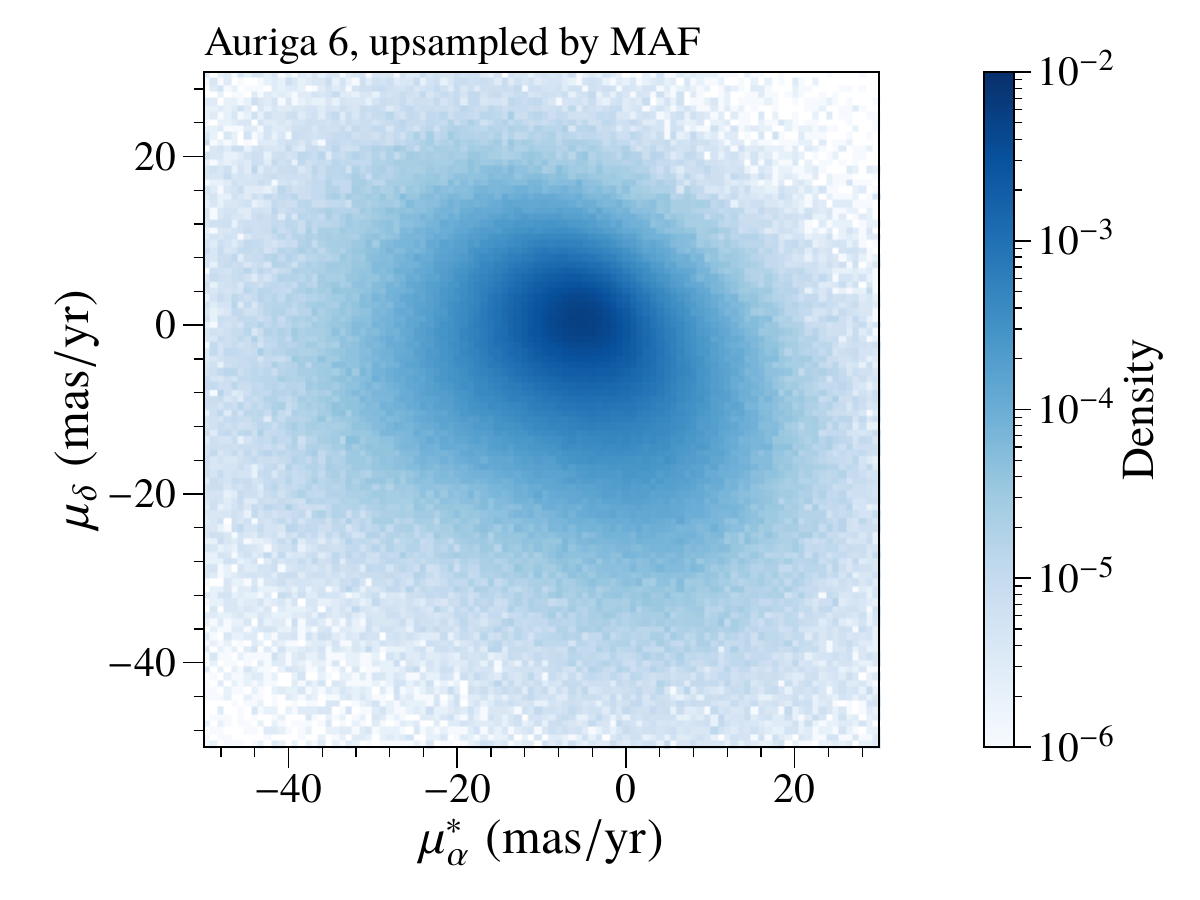}
    \end{center}
    \caption{
        Plots of proper motion for upsampled Auriga~6 stars by MAF within $15^\circ$ of the ICRS coordinate $(\alpha,\delta) = (167.47^\circ,-4.2^\circ)$ with distance from the assumed Solar location within $3.5$~kpc.
    }
    \label{fig:auriga_upsampling_maf}
\end{figure}

\Fig{fig:auriga_upsampling_maf} shows the proper motion distribution of stars upsampled by MAF.
The proper motion distribution is as smooth as the CNF plot in \Fig{fig:auriga_upsampling} by eye, suggesting that the MAF also models the distribution well.

\subsection{Multi-Model Classifier Test: MAF vs.~CNF}

To quantitatively compare the upsampling performance of MAF and CNF, we perform the multi-model classifier test.
The classifier test is performed in the same way described in \sec{sec:classifier_test} but replacing the \enbid{}-upsampled dataset with one from the MAF.

\begin{figure}
    \begin{center}
        \includegraphics[width=0.5\textwidth]{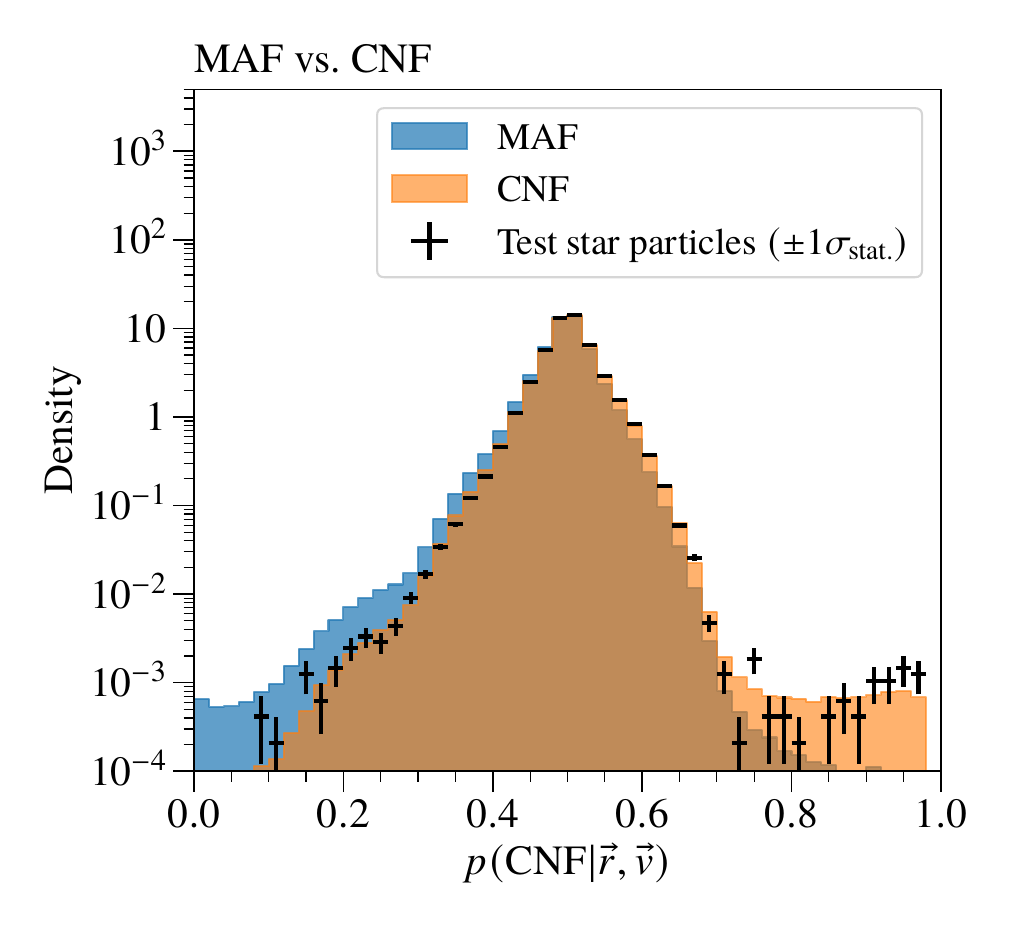}
    \end{center}
    \caption{
        MAF (blue) vs. CNF (orange) classifier output distributions.
        Black markers are the classifier output histogram of test star particles, which is not used in training upsamplers.
        Their vertical error bar is $1\sigma$ statistical uncertainty.
    }
    \label{fig:classifier_maf_cnf}
\end{figure}

We show the classifier output distributions of MAF and CNF-upsampled datasets in \Fig{fig:classifier_maf_cnf}.
Both distributions peak at 0.5, so the two datasets are consistent at some level.
However, the distributions are different in both lower and upper tails, and the AUC of this classifier is $0.5366$, which is slightly higher than 0.5.
This indicates that there are non-negligible differences in the details of the distributions.
The classifier output distribution of the test star particles is close to that of the CNF-upsampled stars, indicating that CNF can be a better upsampler than MAF.

The log-posterior of the classifier test for MAF vs. CNF is shown in the second column of \Tab{table:log_posterior_all}.
We note that the base classifier and shorter training setups give us compatible results.
MAF and CNF are modeling smooth distribution, so the regularization issue due to the spectral bias is not happening here. 
Therefore, we will regard the log-posteriors as estimated log-posteriors, not the log-posterior bounds.
The two scores are much closer than \enbid{} vs. CNF since the classifier output distributions are similar.
Nevertheless, there is a statistically significant difference between them, and CNF shows a better log-posterior.

We conclude that the density estimate learned by the CNF is more accurate than that learned by the MAF. 
Note that MAF is essentially a chain of conditional linear transformations that expand and shrink axes, and small wrinkles may appear during the transformation \citep{pmlr-v80-huang18d}.
Data preprocessing and ensemble averaging may mitigate these effects, but residual artifacts may continue to impact the quality of the density estimation.

\begin{table}
    \caption{
        Log-posteriors from the classifier tests for \enbid{} vs.~MAF vs.~CNF upsampling Auriga 6 star particles.
        The first two columns are the two-model classifier test results for comparing CNF to the other upsampler.
        The last column is the results of the multi-model classifier test comparing \enbid{}, MAF, and CNF all at once. 
        Large (i.e.~less negative) log-posterior scores indicate better upsampling performance.
        We also show the $1\sigma$ standard deviations of the scores obtained by repeating ten times the upsampling and classifier training.
        The bold values are the scores of the best model determined by the test.
    }
    \label{table:log_posterior_all}
    \begin{center}
        \begin{tabular}{cccc}
            \hline
            Classifier: & \enbid{} vs. CNF & MAF vs. CNF & All\\
            \hline
            Upsampler & \multicolumn{3}{c}{$\mathrm{LP}(\mathrm{upsampler})$} \\
            \hline
            \enbid{}  & $-0.724 \pm 0.003$  &                      & $ -1.133 \pm 0.004$ \\
            MAF       &                     & $-0.7021 \pm 0.0016$ & $-1.104 \pm 0.003$ \\
            CNF       & $\mathbf{-0.683 \pm 0.003}$  & $\mathbf{-0.6899 \pm 0.0016}$ & $\mathbf{-1.090 \pm 0.003}$ \\
            \hline
        \end{tabular}
    \end{center}
\end{table}

\subsection{Multi-Model Classifier Test: \enbid{} vs. MAF vs. CNF}

Next, we consider all the upsamplers discussed in this paper -- \enbid{}, MAF, and CNF -- to demonstrate the comparison of multiple upsamplers all at once by our multi-model classifier test.

\begin{figure*}
    \begin{center}
        \includegraphics[width=0.33\textwidth]{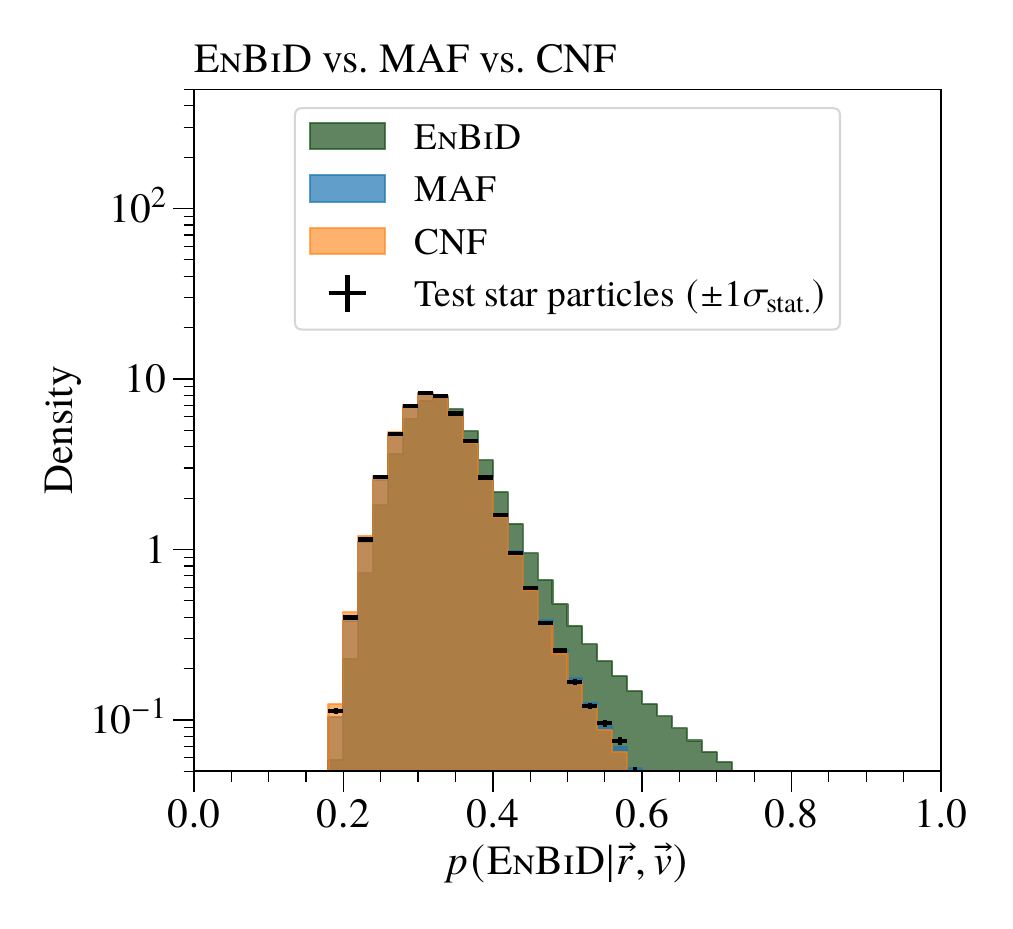}
        \includegraphics[width=0.33\textwidth]{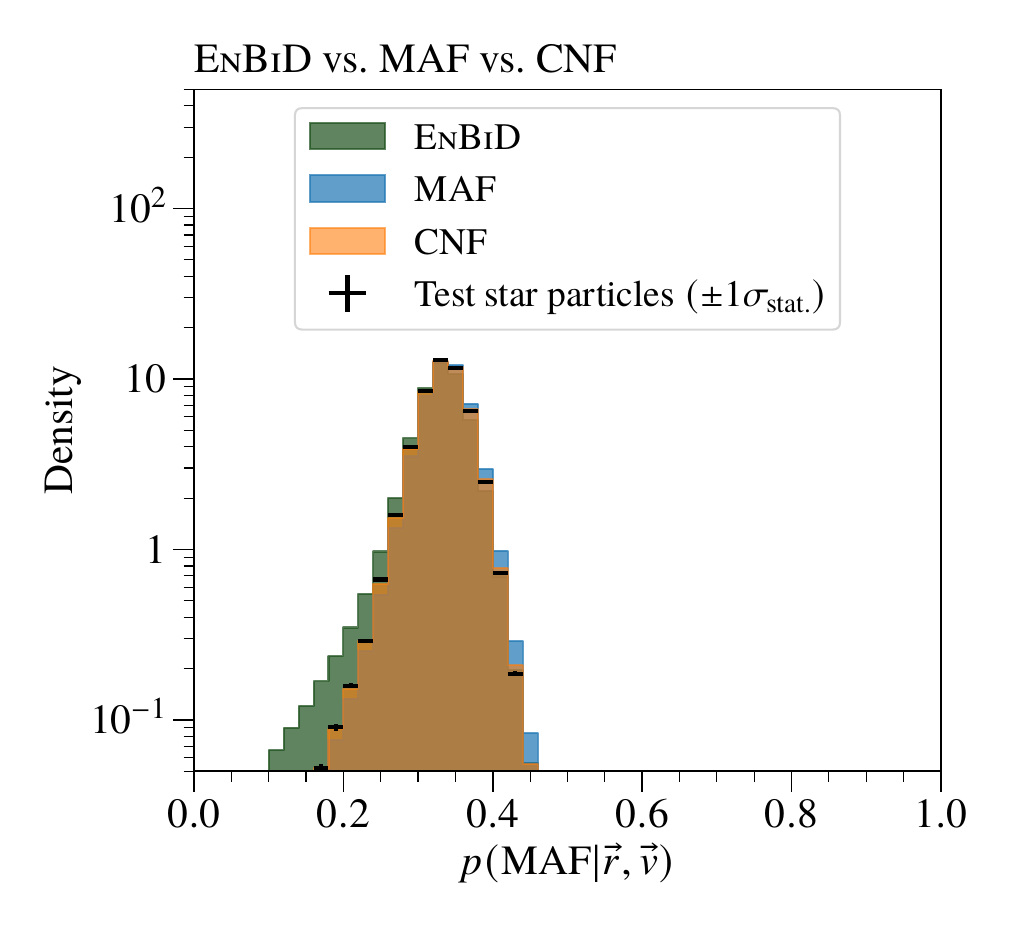}
        \includegraphics[width=0.33\textwidth]{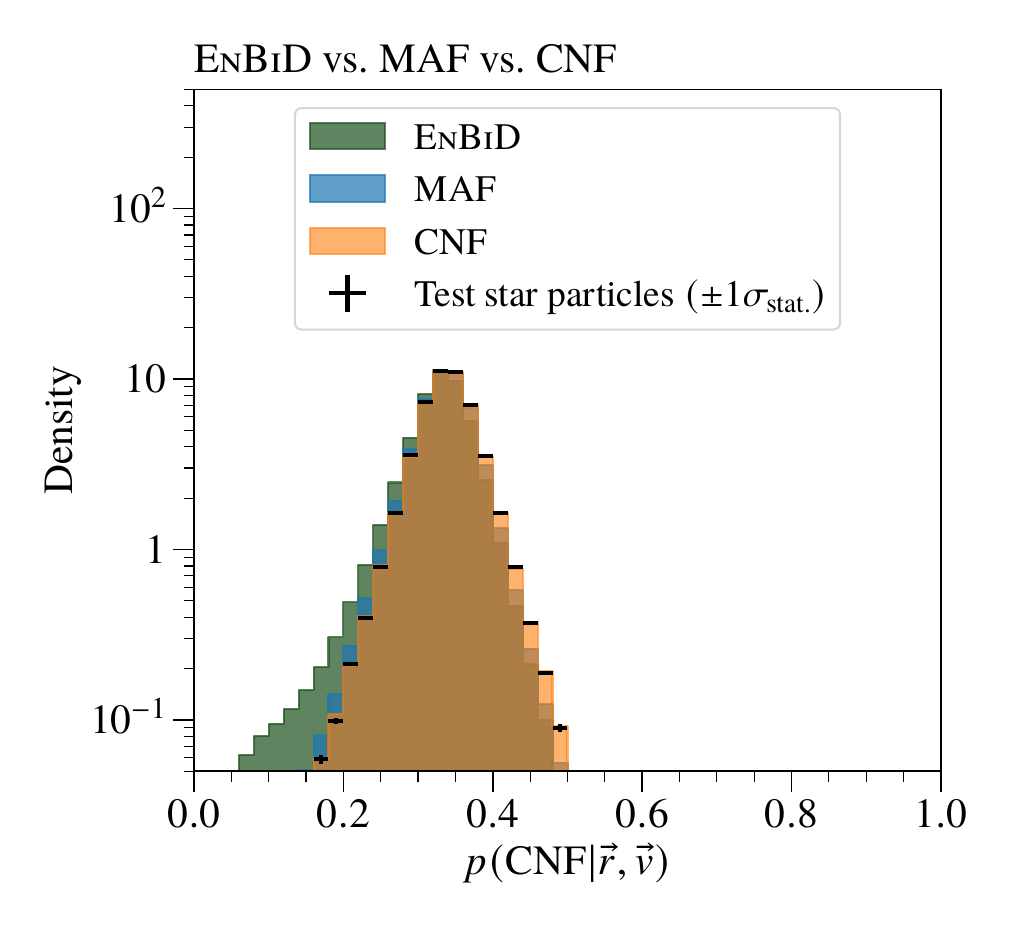}
    \end{center}
    \caption{
        \enbid{} (green) vs. MAF (blue) vs. CNF (orange) classifier output distributions.
        Black markers are the classifier output histogram of test star particles, which is not used in training upsamplers.
        Their vertical error bar is $1\sigma$ statistical uncertainty.
    }
    \label{fig:classifier_enbid_maf_cnf}
\end{figure*}

The classifier output distribution is shown in \Fig{fig:classifier_enbid_maf_cnf}, and we can see the same results we have found from \enbid{} vs. CNF and MAF vs. CNF classifier tests.
First of all, the \enbid{} distribution is clearly different from those of MAF, CNF, and reference dataset, which indicates that \enbid{} is less accurate than the other upsamplers.
The MAF and CNF distributions closely align with the reference dataset, indicating their high accuracy as both models are free from the smoothing bias inherent in kernel density estimations.
Among them, CNF distribution is closest to the reference dataset, suggesting CNF is the most accurate among the upsamplers.

The log-posteriors are listed in the third column of \Tab{table:log_posterior_all}. 
The log-posterior of \enbid{}, MAF, and CNF is $-1.133 \pm 0.004$, $-1.104 \pm 0.003$, and $-1.090 \pm 0.003$, respectively.
Since the log-posterior of CNF is highest, this multi-model classifier test quantitatively concludes that CNF is the best upsampler among those three.

\section{Limitations of MLP-based Multi-Model Classifier Test}
\label{app:justification}

In this appendix, we revisit an overly small bandwidth \enbid{} setup seen in the Ananke and Aurigaia catalogs, which led to blotchiness in the upsampled star distribution, in order to discuss the limitations of using MLP-based classifiers in our multi-model classifier test. 
The main concern is that the inductive biases of MLPs may influence the test outcome if the classifier is underfitted due to the biases.
Therefore, it’s crucial to understand the methods' limitations carefully.

The inductive bias mainly considered here is the spectral bias of MLPs, where the training process tends to favor lower-frequency modes initially \citep{pmlr-v97-rahaman19a, 10.1007/978-3-030-36708-4_22, yang2022overcoming}. 
This bias becomes particularly relevant when dealing with the high-frequency modes of the suboptimal \enbid{}-upsampled star distribution, namely the blotchiness of the dataset at the bandwidth scale. 
We will first discuss the \enbid{} setup that replicates upsampling results similar to those in Aurigaia and examine the effect of the spectral bias on multi-level classifier tests involving such data.

\subsection{\enbid{} Upsampling with Small Bandwidth}

In order to generate blotchy stellar distributions similar to Aurigaia, we adopt the following procedure for upsampling each star particle, similar to the bandwidth determination step described in \citet{2018MNRAS.481.1726G}.
\begin{enumerate}
    \item Use \enbid{} 2.0 package \citep{enbid} to determine the phase space hyperrectangle enclosing the star particle.
    \item For coordinate axis $i$, denote $L_i$ as the corresponding hyperrectangle side length.
    \item Draw phase space coordinates from a 6D Gaussian centered on the original star particle, with each phase space direction's dispersion given by $\sigma_i^2 = \frac{1}{48}(\frac{L_i}{2})^2$.
\end{enumerate}
We select upsampled star particles within a 3.5~kpc radius from the Solar location for consistency with samples generated using normalizing flows.

\begin{figure*}
    \begin{center}
        \includegraphics[height=0.400\textwidth, trim={0 0 4.5cm 0}, clip]{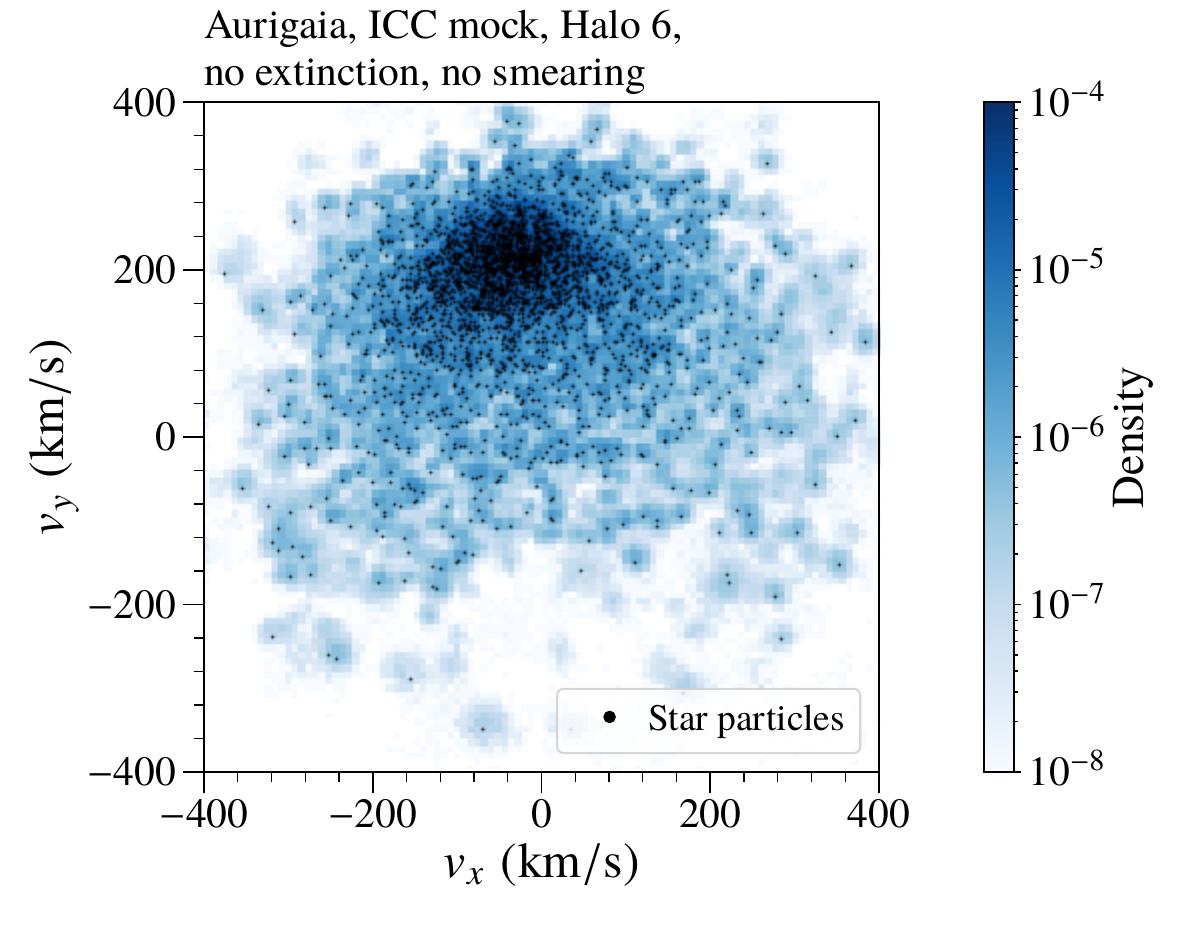}
        \includegraphics[height=0.400\textwidth]{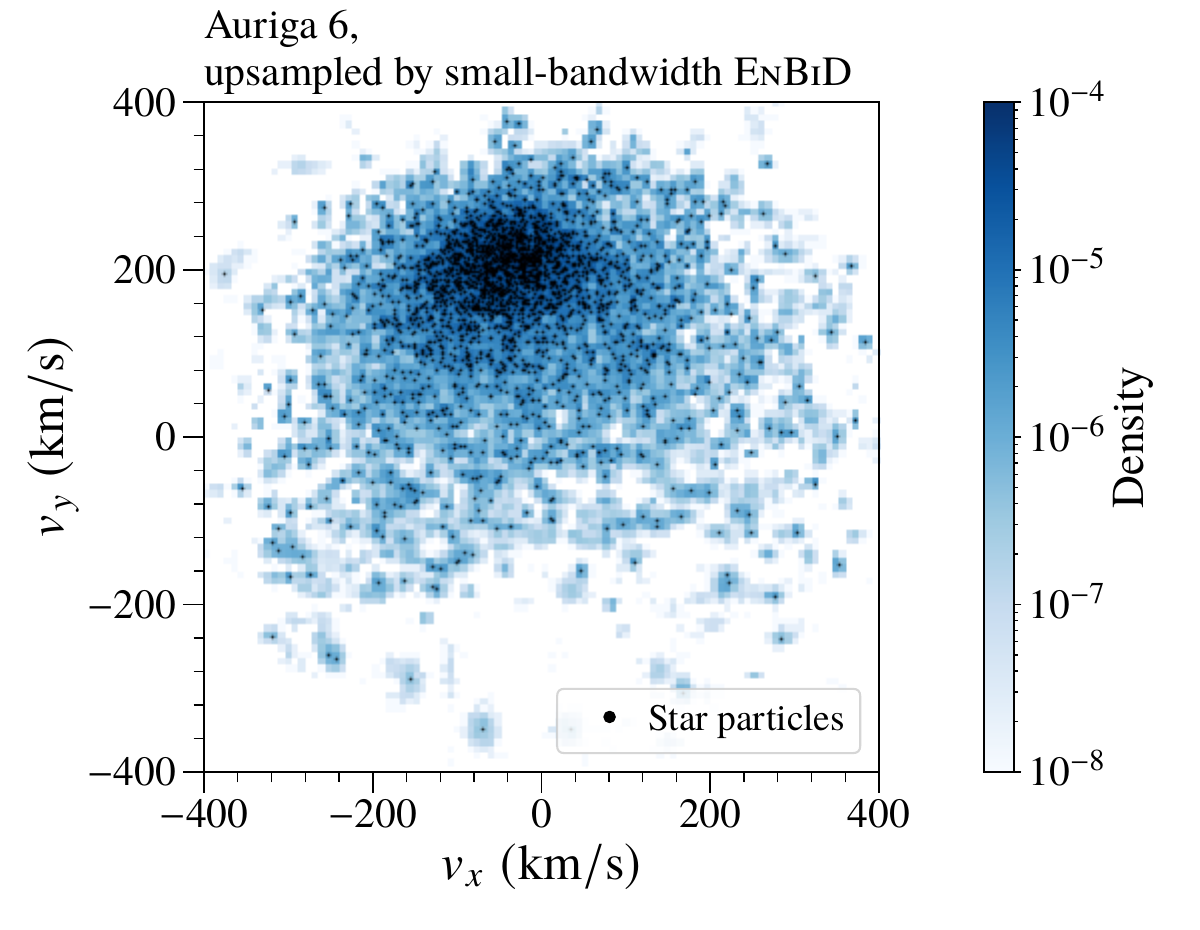}
    \end{center}  		
    \caption{
        Distributions of Galactocentric velocities $(v_x, v_y)$ for Aurigaia (left column) and the Auriga~6 simulation upsampled by \enbid{} (right column). 
        We plot all stars within $15^\circ$ of the ICRS coordinate $(\alpha,\delta) = (167.47^\circ,-4.2^\circ)$ and with distance from the assumed Solar location less than $3.5$~kpc.
        We further select stars with progenitor star particles within 3.5 kpc from the Sun in order for a fair comparison to our \enbid{}-upsampled dataset.
        Black dots are the progenitor star particles.
        The distribution for the Aurigaia catalog is shown without measurement errors.
    }
    \label{fig:aurigaia-no-error}
\end{figure*}

\Fig{fig:aurigaia-no-error} shows the true Galactocentric velocity, without smearing from measurement errors, of stars in the Aurigaia catalog and small-bandwidth \enbid{}-upsampled Auriga~6 within the same sky patch as \Fig{fig:competitor_distributions_mu}.
To ensure fairness in our comparison, stars from the Aurigaia catalog originating from progenitor star particles beyond 3.5 kpc from the Sun are excluded. 
We also overlay the progenitor star particles within this analysis volume for reference.
In each plot, we see the same increased concentration of stars around the star particles, while the stars in the Aurigaia catalog have a larger spread.
The blotches not connected to identified star particles originate from star particles just outside the patch.
Since this appendix focuses on the impact of clumpy substructures in multi-model classifier tests, this small-bandwidth \enbid{} upsampling method is sufficient for our purpose.

\subsection{Multi-Model Classifier Test: Small-Bandwidth \enbid{} vs. CNF}

\begin{figure}
    \includegraphics[width=0.495\textwidth]{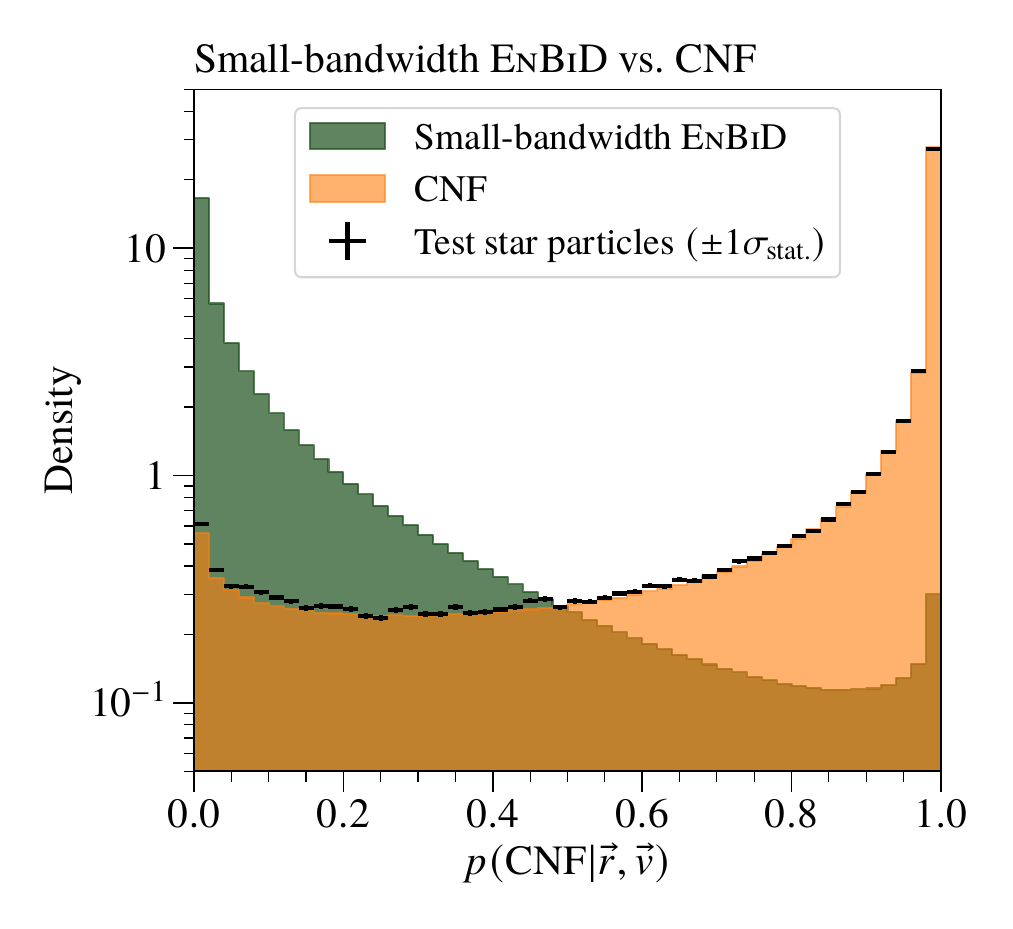}
    \caption{
        Small-bandwidth \enbid{} (green) vs. CNF (orange) classifier output distributions.
        Black markers are the classifier output histogram of this specific feature star particles from the test set, which is not used in training upsamplers.
        Their vertical error bars are the $1\sigma$ statistical uncertainties.
    }
    \label{fig:classifier_enbid_cnf_smallbw}
\end{figure}

\begin{table}
    \caption{
    Log-posteriors from the multi-model classifier test for small-bandwidth \enbid{} vs.~CNF.
    Large (i.e.~less negative) log-posterior scores indicate better upsampling performance.
    We also show the $1\sigma$ standard deviations of the scores obtained by repeating ten times the upsampling and classifier training.
    The bold values are the scores of the best model determined by the test
    }
    \begin{center}
        \begin{tblr}{cc}
            \hline
            Classifier: & \enbid{} vs.~CNF
            \\
            \hline
            Upsampler & $\mathrm{LP}(\mathrm{upsampler})$
            \\
            \hline
            \enbid{}  & $<-2.48 \pm 0.19$  \\
            CNF       & $\mathbf{>-0.54 \pm 0.01}$  \\
            \hline
        \end{tblr}
    \end{center}
\label{table:log_posterior_enbid_cnf_sub}
\end{table}

Here, we present the multi-model classifier test results comparing small-bandwidth \enbid{} and CNF upsampling method.

\Fig{fig:classifier_enbid_cnf_smallbw} shows the classifier output distributions for both upsampling methods, and the two distributions are well separated. 
Stars exhibiting high CNF posterior probability $p(\mathrm{CNF}|\vec{r},\vec{v})$ are mostly located in the gaps between \enbid{}'s narrow Gaussian kernels as the likelihood of being \enbid{}-upsampled is low.
Meanwhile, stars located atop the kernels have a high likelihood of being \enbid{}-upsampled, so that the posterior probability will be smaller.
As a result, the two posterior probability distributions are clearly separated. 
The AUC of this classifier is 0.952, confirming that the classifier effectively distinguishes the two types of upsampled stars.

Next, we use the classifier to determine which upsampler generates stars whose distribution most closely resembles that of the original star particles.
Considering that the test star particles are mostly positioned away from the \enbid{}'s Gaussian kernels, we expect that the classifier will recognize the test star particles as CNF-upsampled stars.
Indeed, the classifier output distribution of the test star particles is aligned to that of the CNF-upsampled dataset, as shown in \Fig{fig:classifier_enbid_cnf_smallbw}. 
This alignment suggests that CNF more accurately replicates the original star particle distribution.

Although the results of this multi-model classifier test align with our qualitative understanding, we cannot use this classifier for precise log-posterior estimation, primarily due to the signs of underfitting: the training continues if we increase the patience of early stopping.
We may further continue training to mitigate this issue; however, the training is too slow.
The base classifier training setup finishes in 6 days using a Tesla P100 GPU already, and further pushing training is not feasible with our available computational resources.

Furthermore, this slow training speed also complicates the uncertainty estimation,
Ideally, the training should be repeated several times with different random number seeds in order to assess the variance of the log posterior.
However, the extensive computational time required for each iteration makes this uncertainty estimation impractical in this case.

This slow training pace is due to the spectral bias of MLPs \citep{pmlr-v97-rahaman19a, 10.1007/978-3-030-36708-4_22, yang2022overcoming}: training of neural networks tends to prioritize learning lower frequency modes first.
In the above classifier training, the early stopping effectively acts as a regularizer since the spectral bias significantly slows down the learning of features on high frequencies, such as the clumpy substructures of \enbid{} upsampled dataset.
The early stopping prematurely terminates the training before the neural network learns all the relevant high-frequency features.
Thus, rather than accurately regressing the \enbid{}-upsampled distribution, the classifier is constrained to learning a partially smoothed version of the distribution due to this implicit regularization.

Instead of continuing the training forever to get more accurate log posteriors, we admit the underfitting and perform the multi-model classifier test with a shorter classifier training setup by making the mini-batch size 10 times larger.
The effective patience for network parameter updates is then one-tenth of the base setup, so the early stopping will terminate the training earlier than the base setup.
This regularized classifier can still distinguish the small-bandwidth \enbid{} and CNF (AUC: 0.831), and the training finishes within 10 hours.
The decrease in AUC is as expected since the regularized classifier is no longer a linear function of the log-likelihood ratio and optimal classifier.

We show the log-posteriors of the multi-model classifier test using this regularized classifier in \Tab{table:log_posterior_enbid_cnf_sub}.
Specifically, the log-posterior scores for \enbid{} and CNF are $-2.48\pm0.19$ and $-0.54\pm0.01$, respectively.
We interpret these log-posteriors from the regularized classifier as indicative bounds -- upper for \enbid{} and lower for CNF -- relative to their asymptotic values because the base classifier gives us smaller and larger log-posterior, respectively.
The log-posterior from the base classifier $-6.22$ for \enbid{}, and $-0.34$ for CNF.
The quoted uncertainties on the log-posterior scores are derived by repeating the test ten times with different random number seeds.
This uncertainty covers the statistical uncertainty of repeated upsampling and the systematic uncertainty of network initialization.

Although a regularized classifier was employed in this analysis, the difference between the two log-posteriors is nonzero and statistically significant. 
This means the classifier, when asked whether the dataset of test star particles looked more like the CNF-upsampled data or the \enbid{}-upsampled data, determined that it had a higher likelihood of being drawn from the CNF-upsampled dataset than \enbid{}-upsampled dataset.
Therefore, this test result still quantitatively demonstrates that our \GalaxyFlow{} algorithm based on a CNF generates stars more kinematically consistent with the underlying star particle distribution.

Note that this limitation of the MLP-based multi-level classifier test can be avoided if we use a classifier that does not exhibit spectral bias. 
While classifiers free from spectral bias could potentially offer more accurate log-posterior estimation, introducing such classifiers here falls beyond the scope of the paper.

\section{Results on Galaxy h277} 
\label{app:h277}

In this appendix, we apply our upsampling method to the galaxy h277\footnote{The simulation data for this galaxy is available from the $N$-Body Shop Collaboration at \url{https://nbody.shop/data.html}.} \citep{2012ApJ...761...71Z,2012ApJ...758L..23L} from the N-Body Shop Collaboration.
The galaxy h277 was produced using the Gasoline smoothed particle hydrodynamics code \citep{2004NewA....9..137W}, starting with a dark matter-only simulation within a comoving box with side-length 50~Mpc, and then re-simulating with identical initial conditions at higher resolution including baryons. This simulation has a force resolution of 173~pc, dark matter particle mass of $1.3\times 10^5\,M_\odot$, initial gas particle mass $2.7\times 10^4\,M_\odot$, and average star particle mass of $5800\,M_\odot$. The entire galaxy has a mass of $7.95 \times 10^{11}\, M_\odot$ within the virial radius, and was selected as the most Milky Way-like of the galaxies available from this set of simulations.
This galaxy currently has no associated mock \Gaia{} catalog, so we here present our upsampler as a first step towards such a catalog.

The star particles within 3.5 kpc from the assumed Solar location, $(-8.122, 0.0, 0.0208)$ kpc,\footnote{We use \textsc{Pynbody} \citep{pynbody} to orient the Cartesian Galactocentric coordinate system of h277 so that the galactic disks lie in the $xy$-plane. 
We further rotate the frame about the $z$ axis by $-20^\circ$ and flip the $x$ axis.} are selected for training our upsampler.
The number of selected star particles is 153,724, and the maximum speed of the selected star particles is 433.79 km/s.
We center and re-scale our particles and train a CNF following the procedures outlined in \sec{sec:nfs}.

\begin{figure*}
    \centering
    \includegraphics[width=0.95\textwidth]{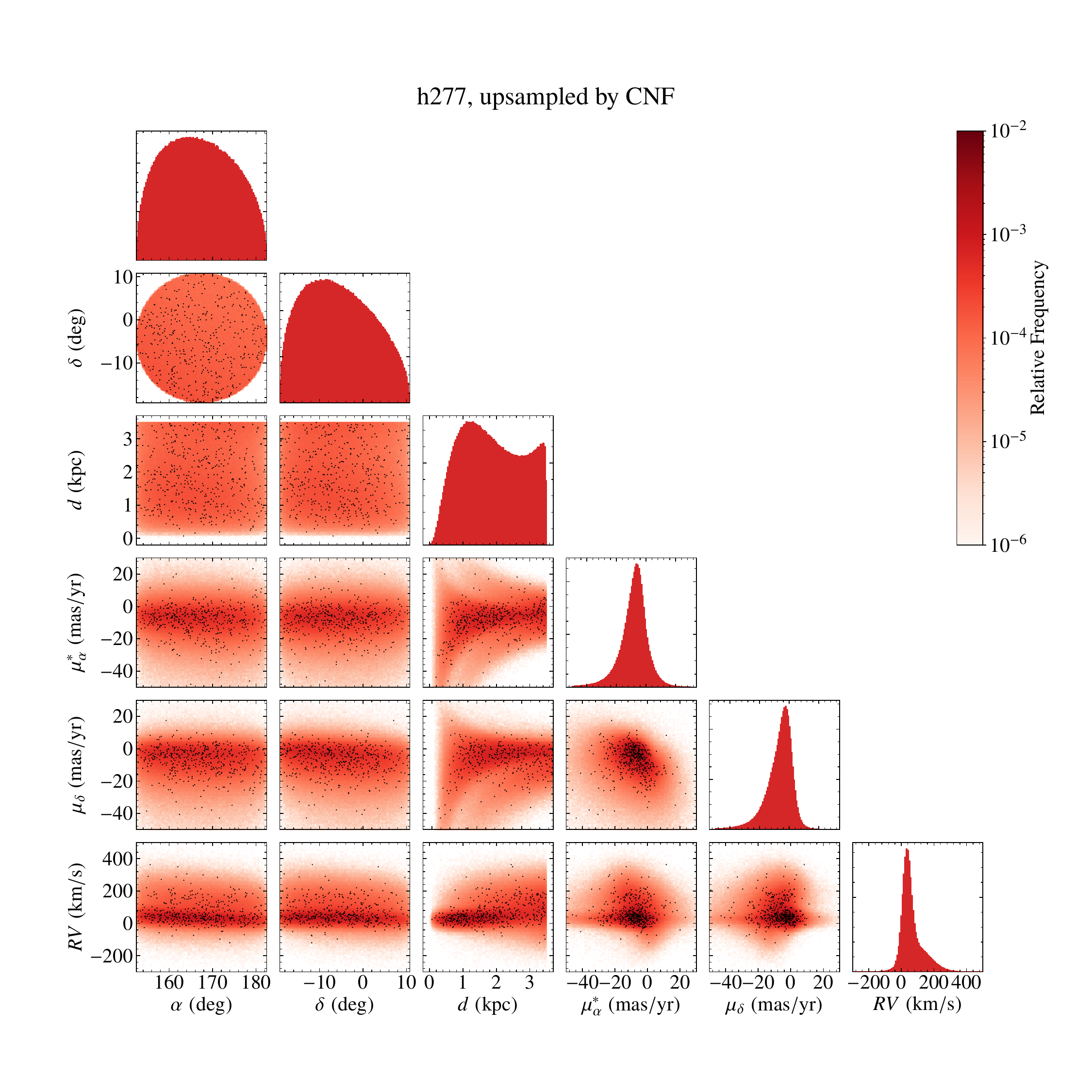}
    \caption{
        Distributions of CNF-upsampled h277 stars in the ICRS coordinate.
        We select stars within $15^\circ$ from $(\alpha,\delta) = (167.47^\circ,-4.2^\circ)$ and with distances from the assumed Solar location less than $3.5$~kpc.
        The star particles within the selected region are shown as black dots.
        This patch contains 440 star particles. 
  	}
    \label{fig:h277_icrs}
\end{figure*}

In \Fig{fig:h277_icrs}, we show the phase space density distributions in ICRS coordinates for stars upsampled in the angular patch as \Fig{fig:competitor_distributions_mu}.
We additionally overlay the progenitor star particles on each 2D distribution.
The patch contains 440 star particles, upsampled to 2,429,709 stars — approximately 5,800 stars per particle or 1 star per $M_\odot$.
The distributions have no star particle substructure, and the CNF learns the bimodal distance distribution.
These plots indicate that upsampling with normalizing flows provides consistent results across multiple simulations.

\begin{table}
    \caption{
        Log-posteriors from the classifier tests for \enbid{} vs.~CNF upsampling h277 star particles.
        Large (i.e.~less negative) log-posterior scores indicate better upsampling performance.
        We also show the $1\sigma$ standard deviations of the scores obtained by repeating ten times the upsampling and classifier training.
        The bold values are the scores of the best model determined by the test.
    }
    \label{table:log_posterior_all:h277}
    \begin{center}
        \begin{tabular}{ccc}
            \hline
            Classifier: & \enbid{} vs.~CNF
            \\
            \hline
            Upsampler & $\mathrm{LP}(\mathrm{upsampler})$
            \\
            \hline
            \enbid{}  &  $ -0.736 \pm 0.003$  \\
            CNF       &  $\mathbf{-0.681 \pm 0.003}$  \\
            \hline
        \end{tabular}
    \end{center}
\end{table}

We then performed multi-model classifier tests for comparing \enbid{} and CNF-generated datasets. 
For \enbid{}, we find that the optimal bandwidth multiplier is $0.3348 \pm 0.0016$, which is compatible with the theoretical estimation from the rule-of-thumb bandwidth ratio, 0.332.
The log-posteriors of \enbid{} and CNF are shown in \Tab{table:log_posterior_all:h277}.
Again, the \enbid{}-upsampled dataset contains kernel smoothing bias, and the log-posterior of \enbid{} is small.
The log-posterior for CNF is larger than that, and these results again confirm that our \GalaxyFlow{} using CNF is more accurate than \enbid{} upsampling.

\section{Kernel Bandwidth Scaling} 
\label{app:rot_bandwidth}

In this appendix, we discuss the bandwidth scaling of kernel density estimation in order to translate the optimal bandwidth for one kernel to another.
We first discuss the bandwidth scaling using the rule-of-thumb bandwidths of kernel density estimation for $D$-dimensional data, and then generalize the derivation for adaptive kernel density estimations, such as the \enbid{} method.
For a more general review of the rule-of-thumb bandwidths, we refer to the following textbooks: \citet{econ, doi:https://doi.org/10.1002/9780470316849.ch6}.

\subsection{Rule-of-Thumb Bandwidth Ratio}
The rule-of-thumb bandwidth of kernel density estimation (KDE) is an asymptotically\footnote{Large number of samples and small bandwidth limit.} optimal bandwidth when a given dataset is approximately Gaussian distributed.
When the dataset is not too complicated, it is often considered an initial bandwidth choice.
In this subsection, we derive the rule-of-thumb bandwidth and discuss the use of their ratio for the bandwidth scaling.

For simplicity of discussion, we will only consider spherically symmetric kernels $K$, i.e., kernels that are only a function of the magnitude of inputs.
For example, a Gaussian kernel $K^G$ and a spherical Epanechnikov kernel $K^E$ are spherically symmetric kernels defined as follows:
\begin{eqnarray}
    K^G(x^a) 
    & = &
    \frac{1}{(2 \pi)^{D/2}} e^{-\frac{||x^a||^2}{2}},
    \\
    K^E(x^a) 
    & = & 
    \begin{cases}
        \frac{\Gamma\left[2+\frac{D}{2}\right]}{\pi^{\frac{D}{2}}}(1-||x^a||^2) & ||x^a|| < 1, \\
        0 & ||x^a|| > 1,
    \end{cases}
\end{eqnarray}
where $x^a$ is a $D$-dimensional input vector with $a=1,\cdots,D$.
We denote the scaled kernel of $K$ with a bandwidth $h$ as 
\begin{equation}
    K_h(x^a) = \frac{1}{h^D} \cdot K\left(x^a/h\right).
\end{equation}

Kernel density estimation $\hat{f}$ is then the average of scaled kernels with a fixed bandwidth $h$ and centered on each of the data points, i.e.,
\begin{equation}
    \hat{f}(x^a; X^a_{(i)}) = \frac{1}{N} \sum_{i=1}^N K_{h}(x^a - X^a_{(i)}), 
\end{equation}
where $X^a_{(i)}$ is the $i$-th sample drawn from the true density $f(x^a)$. 
The expectation of $\hat{f}$ is simply the convolution of $f$ and $K_h$,
\begin{equation}
    E_X  \left[ \hat{f}(x^a; X^a_{(i)})  \right]
    = 
    \frac{1}{N} \sum_{i=1}^N \int d^D x' f (x'^a) K_{h}(x^a - x'^a) = (f * K_h)(x^a).
\end{equation}
The bias and variance of $\hat{f}$ is then as follows:
\begin{eqnarray}
    \mathrm{Bias}(\hat{f})(x^a) 
    & = &
    E_X \left[
        \hat{f}(x^a; X^a_{(i)}) - f(x^a)
    \right]
    \\
    \label{eqn:bias_kde}
    & = &
    (f * K_h)(x^a) - f(x^a),
    \\
    \mathrm{Var}(\hat{f})(x^a) 
    & = & 
    \frac{1}{N^2} \sum_{i=1}^N \mathrm{Var}\left[ 
        K_h (x^a - X^a_{(i)} )\right]
    \\
    & = &
    \frac{1}{N} \left[ (f * K_h^2)(x^a) - (f*K_h)(x^a) \right].
\end{eqnarray}
Note that $\hat{f}$ is a biased estimator of the true density $f$ since the bias \eqref{eqn:bias_kde} is nonzero.

The rule-of-thumb bandwidth is then derived from an analytic solution of minimizing mean integrated square error (MISE) defined by an integral of mean square error (MSE).
\begin{eqnarray}
    \mathrm{MISE} 
    & = & 
    \int d^D x \, \mathrm{MSE}(x^a). 
    \\
    \mathrm{MSE}(x^a) 
    & = &
    E_{X}\left[ (\hat{f}(x^a; X^a_{(i)}) - f(x^a))^2 \right],
    \\
    & = &
    \mathrm{Var}(\hat{f})(x^a) + \left[  \mathrm{Bias}(\hat{f})(x^a)  \right]^2.
\end{eqnarray}

In order to discuss the asymptotic behavior of MISE, we Taylor expand the bias and variance of $\hat{f}$.
The expansion of bias is as follows:
\begin{equation}
    \mathrm{Bias}(\hat{f}) = \frac{1}{2} h^2 \mu_2(K) \sum_{a=1}^D \frac{d^2 f}{d (x^a)^2} + \cdots.
\end{equation}
The leading term of this expansion is proportional to the second spherical moment of the kernel, $\mu_2(K)$,
\begin{equation}
    \mu_2(K) = \frac{1}{D} \int d^D ||x^a||^2 K(x).
\end{equation}
The Taylor expansion of variance is as follows:
\begin{equation}
    \mathrm{Var}(\hat{f}) = 
        \frac{f(x^a) R(K)}{N h^D} + \cdots.
\end{equation}
Here, $R(K)$ is called a roughness of kernel,
\begin{equation}
    R(K) = \int d^D x \, K^2(x).
\end{equation}

Now, let us try to find a bandwidth solution minimizing the asymptotic expression of MISE. 
The sum of the dominant terms in the asymptotic expansion of MISE is called asymptotic MISE, 
\begin{equation}
    \label{eqn:kde:amise}
    \mathrm{AMISE} = \frac{1}{4}  h^4 \mu_2(K)^2 R\left[ \nabla^2 f \right] + \frac{R(K)}{N h^D}.
\end{equation}
The minimization of AMISE is analytically solvable, and the solution bandwidth is as follows:
\begin{equation}
    h(K) = \left[ \frac{1}{N} \frac{D R(K)}{ \mu_2(K)^2 R\left[ \nabla^2 f \right]} \right]^{D+4}.
\end{equation}
There is a data-dependent roughness term $R\left[ \nabla^2 f \right]$, so the asymptotic solution depends on data in general. 
Nevertheless, if the data distribution is close to Gaussian, we may replace $R\left[ \nabla^2 f \right]$ with that of Gaussian.
The resulting bandwidth is called the rule-of-thumb bandwidth for a kernel $K$.

In the case of considering ratios of the optimal bandwidth between two kernels, the discussion is much simpler since the roughness term $R\left[ \nabla^2 f \right]$ cancels out. 
The ratio is then simply a function of the second moment and the roughness of two kernels,
\begin{equation}
    \label{eqn:rule_ratio}
    \frac{h(K^G)}{h(K^E)} = \left[ \frac{\mu_2(K^E)^2}{\mu_2(K^G)^2} \frac{R(K^G)}{R(K^E)} \right]^{\frac{1}{D+4}}.
\end{equation}
The second moment and the roughness of Gaussian and spherical Epanechnikov kernels in 6 dimensions are as follows:
\begin{align}
\mu_2(K^G) & = 1, & 
R(K^G) & = \frac{1}{64 \pi^3}, \\
\mu_2(K^E) & = \frac{1}{10}, & 
R(K^E) & = \frac{48}{5 \pi^3}. \\
\end{align}
If we plug these numbers into \eqref{eqn:rule_ratio}, we get the scaling factor converting bandwidth for the spherical Epanechnikov kernel to that for the Gaussian kernel, 
\begin{equation}
     \frac{h(K^G)}{h(K^E)} = 2^{-\frac{6}{5}} \cdot 15 ^{-\frac{1}{10}} \approx 0.332.
\end{equation}

\subsection{Bandwidth Scaling for Adaptive Kernel Density Estimations}

In this subsection, we extend the above derivation of bandwidth scaling to adaptive kernel density estimations.
For simplicity of discussion, let us assume that the variable bandwidth vector $h^a$ is a function of a corresponding sample $X^a_{(i)}$ and the total number of samples $N$.
This setup is also known as a sample-point adaptive estimator, an example of which is Abramson's adaptive bandwidths \citep{10.1214/aos/1176345986}.
Although the bandwidth given by the \enbid{} algorithm is a function of the entire dataset $\{X^a_{(i)}|i=1,\cdots,N\}$, this is a reasonable approximation of the setup. 
The algorithm utilizes the volume occupied by a sample to compute the bandwidth, and the expectation of this volume is approximately a function of the true density at a given location and the number of samples.

For the following discussions, let $h^a$ be a variable bandwidth given by \enbid{}, and let $c_h$ be the overall coefficient to fine-tune the bandwidths. 
The scaled kernel with a bandwidth vector $c_h h^a$ is then defined as follows:
\begin{equation}
    K_{c_h h^a}(x^a) = \frac{1}{\prod_{a=1}^D c_h h^a} K\left(\frac{x^a}{c_h  h^a}\right).
\end{equation}
The adaptive kernel density estimation $\hat{f}$ is again the sum of the scaled kernels given samples and bandwidths,
\begin{equation}
    \hat{f}(\vec{x}; X^a_{(i)}) = \frac{1}{N} \sum_{i=1}^N K_{c_h h^a_{(i)}}(x^a - X^a_{(i)}).
\end{equation}
The bias and variance of this $\hat{f}$ is given as follows by generalizing the theorem 2 of \citet{10.1214/aos/1176348768}.
\begin{equation}
    \label{eqn:akde:bias}
    \mathrm{Bias}(\hat{f}) = \frac{1}{2} c_h^2 \mu_2(K)
    \cdot 
    \sum_{a=1}^D
        \frac{d^2 }{d (x^a)^2} \left[ f \cdot (h^a)^2 \right] 
    + \cdots . 
\end{equation}
\begin{equation}
    \label{eqn:akde:var}
    \mathrm{Var}(\hat{f}) = 
    \frac{R(K)}{N c_h^D} \cdot \frac{f(x^a)}{\prod_{a=1}^D h^a}  + \cdots .
\end{equation}
We note that the first terms of the above two equations are data-independent while the second terms are not.

The asymptotic MISE of $\hat{f}$ has a form similar to \eqref{eqn:kde:amise},
\begin{equation}
    \mathrm{AMISE} = \frac{1}{4}  c_h^4 \mu_2(K)^2 \cdot C_1(f, h^a) + \frac{R(K)}{N c_h^D} C_2(f, h^a),
\end{equation}
where $C_1$ and $C_2$ are data-dependent functions representing the integrals related to data-dependent terms in \eqref{eqn:akde:bias} and \eqref{eqn:akde:var}, respectively.

Again, the minimization of this AMISE with respect to $c_h$ is analytically solvable, and the data-independent term of the scaling factor is as follows:
\begin{equation}
    c_h(K) \propto \left[ \frac{1}{N} \frac{D R(K)}{ \mu_2(K)^2 } \right]^{D+4}.
\end{equation}
Therefore, the scaling factor ratio is the same as the rule-of-thumb bandwidth ratio in \eqref{eqn:rule_ratio},
\begin{equation}
    \frac{c_h(K^G)}{c_h(K^E)} = \left[ \frac{\mu_2(K^E)^2}{\mu_2(K^G)^2} \frac{R(K^G)}{R(K^E)} \right]^{\frac{1}{D+4}}. 
\end{equation}

\bsp	
\label{lastpage}
\end{document}